%% file: main.tex
\documentclass[lettersize,journal]{IEEEtran}
\usepackage{amsmath,amsfonts}
\usepackage{algorithm}
\usepackage{array}
\usepackage[caption=false,font=normalsize,labelfont=sf,textfont=sf]{subfig}
\usepackage{textcomp}
\usepackage{stfloats}
\usepackage{url}
\usepackage{verbatim}
\usepackage{graphicx}
\usepackage{cite}
\usepackage{wrapfig}
\usepackage[utf8]{inputenc}
\usepackage{colortbl}
\usepackage{xcolor}
\usepackage[export]{adjustbox} 
\usepackage{makecell}
\usepackage{tabularx}
\usepackage{glossaries}
\glsdisablehyper
\usepackage{mathtools}
\PassOptionsToPackage{achronym}{glossaries}
\usepackage{algpseudocode}
\usepackage{subfig}
\DeclarePairedDelimiter{\ceil}{\lceil}{\rceil}

\newacronym{3d}{3D}{three dimensional}
\newacronym{6g}{6G}{6th generation of wireless networks}
\newacronym{ris}{RIS}{reconfigurable intelligent surface}
\newacronym{rx}{RX}{receiver}
\newacronym{tx}{TX}{transmitter}
\newacronym{bs}{BS}{base station}
\newacronym{em}{EM}{electromagnetic}
\newacronym{e2e}{E2E}{end-to-end}
\newacronym{ue}{UE}{user equipment}
\newacronym{snr}{SNR}{signal-to-noise ratio}
\newacronym{v2v}{V2V}{vehicle-to-vehicle}
\newacronym{ula}{ULA}{uniform linear array}
\newacronym{upa}{UPA}{uniform planar array}
\newacronym{pcb}{PCB}{Polychlorierte Biphenyle}
\newacronym{fss}{FSS}{frequency selective surface}
\newacronym{pbc}{PBC}{periodic boundary conditions}
\newacronym{lc}{LC}{inductor-capacitor}
\newacronym{dda}{DDA}{discrete dipole approximation}
\newacronym{cma}{CMSA}{canonical minimum scattering approximation}
\newacronym{mom}{MoM}{method of moments}
\newacronym{sota}{SotA}{state-of-the-art}
\newacronym{hpbw}{HPBW}{half-power beamwidth}

\input{Definitions.tex}
\newtheorem{problem}{Problem}

\newcommand{\change}[1]{#1}

\newcommand{\name}{T3DRIS} 
\newcommand{\rmRIS}{\scriptscriptstyle\mathrm{RIS}}

\newcommand{\rmT}{\scriptscriptstyle\mathrm{T}}
\newcommand{\rmR}{\scriptscriptstyle\mathrm{R}}
\newcommand{\rmS}{\scriptscriptstyle\mathrm{S}}

\newcommand{\rmL}{\scriptscriptstyle\mathrm{L}}

\newcommand{\rmG}{\scriptscriptstyle\mathrm{G}}
\newcommand\thefontsize{\expandafter\string\the\font}
\begin{document}

\title{\name{}: Advancing Conformal RIS Design through\\ In-depth Analysis of Mutual Coupling Effects}

\author{Placido~Mursia,~\IEEEmembership{Member,~IEEE,}
Francesco~Devoti,~\IEEEmembership{Member,~IEEE,}
Marco~Rossanese,~\IEEEmembership{Student Member,~IEEE,}
Vincenzo~Sciancalepore,~\IEEEmembership{Senior Member,~IEEE,}
Gabriele~Gradoni, \IEEEmembership{Senior Member,~IEEE,}
Marco~Di~Renzo,~\IEEEmembership{Fellow,~IEEE,}
Xavier~Costa-P\'erez,~\IEEEmembership{Senior Member,~IEEE}

\thanks{\textit{\hspace{-0.3cm}P. Mursia, F. Devoti, M. Rossanese and V. Sciancalepore, are with NEC Laboratories Europe, 69115 Heidelberg, Germany (email: name.surname@neclab.eu). \newline
G. Gradoni is with University of Surrey, Stag Hill, University Campus, Guildford GU2 7XH, UK (email: g.gradoni@surrey.ac.uk).\newline
M. Di Renzo is with Universit\'e Paris-Saclay, CNRS, CentraleSup\'elec, Laboratoire des Signaux et Syst\`emes, 91190 Gif-sur-Yvette, France (email: marco.di-renzo@universite-paris-saclay.fr). \newline
X. Costa-P\'erez is with i2cat, ICREA, 08034 Barcelona, Spain and NEC Laboratories Europe (email: xavier.costa@neclab.eu). \newline
Email of the corresponding author: placido.mursia@neclab.eu.}}%
}

\markboth{Submitted to IEEE Transactions on Communications}%
{Shell \MakeLowercase{\textit{et al.}}: A Sample Article Using IEEEtran.cls for IEEE Journals}


\maketitle

\begin{abstract}
This paper presents a theoretical and mathematical framework for the design of a conformal \gls{ris} that adapts to non-planar geometries, which is a critical advancement for the deployment of RIS on non-planar and irregular surfaces as envisioned in smart radio environments. Previous research focused mainly on the optimization of \glspl{ris} assuming a predetermined shape, while neglecting the intricate interplay between shape optimization, phase optimization, and mutual coupling effects. Our contribution, the \name{} framework, addresses this fundamental problem by integrating the configuration and shape optimization of \glspl{ris} into a unified model and design framework
, thus facilitating the application of \gls{ris} technology to a wider spectrum of environmental objects. The mathematical core of \name{} is rooted in optimizing the 3D deployment of the unit cells and tuning circuits, aiming at 
maximizing the communication performance. Through rigorous full-wave simulations and a comprehensive set of numerical analyses, we validate the proposed approach and demonstrate its superior performance and applicability over contemporary designs. This study---the first of its kind---paves the way for a new direction in \gls{ris} research, emphasizing the importance of a theoretical and mathematical perspective in tackling the challenges of conformal \glspl{ris}. 
\end{abstract}

\glsresetall
\section{Introduction}
\label{sec:intro}

\subsection{Background \& Motivation}

In the landscape of modern wireless communication systems, the advent of \gls{ris} marks a significant milestone~\cite{DiRenzo2020_Jsac}, introducing the capability to dynamically manipulate the propagation of \gls{em} waves. This property, together with its flexibility in deployment and reconfiguration, low implementation cost and energy consumption, positions \gls{ris} as a primary technology candidate for the upcoming \gls{6g}~\cite{long2021promising}, opening up the possibility of optimizing the communication channel to best serve different use cases that \gls{6g} will support~\cite{renzo2019smart, basharat2021reconfigurable}. Conventional implementations have predominantly explored planar \gls{ris} architectures, which, while effective, encounter limitations in adaptability and versatility, particularly when confronted with the complexity of non-planar deployment scenarios~\cite{Cui2023}. 

\begin{figure}
    \centering
    \includegraphics[width=0.95\linewidth]{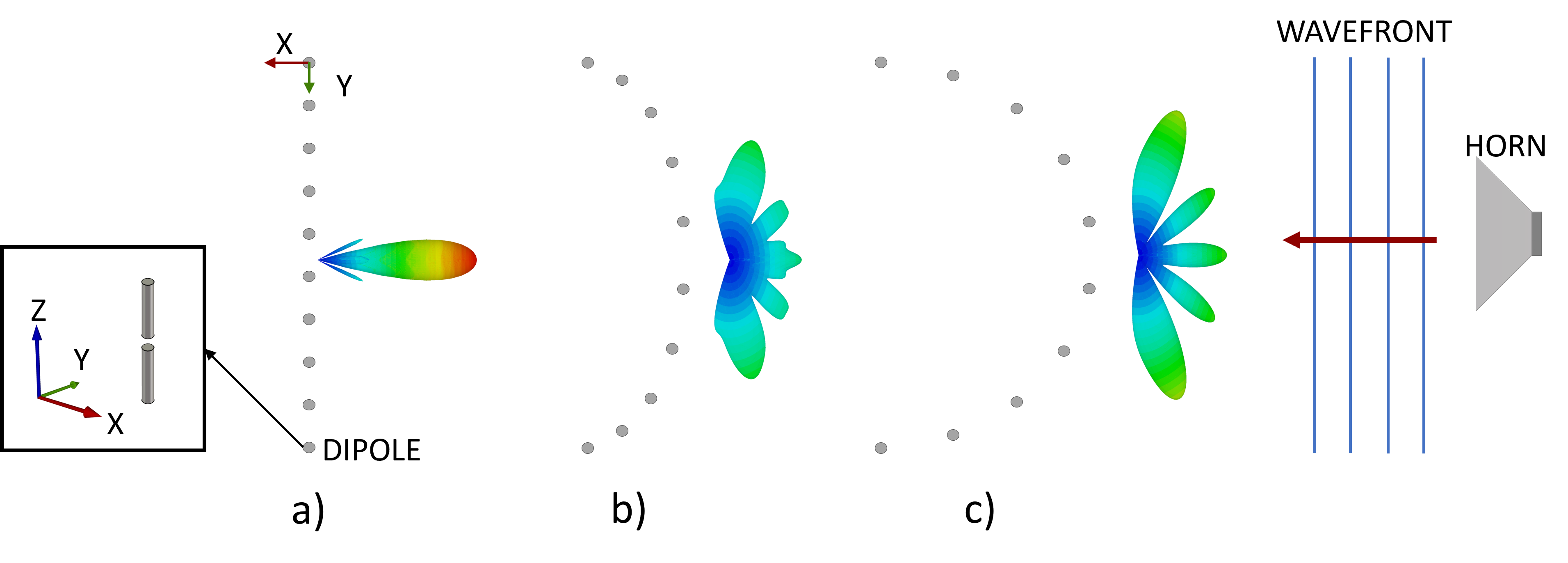}
    \caption{Reflection properties of curved surfaces ($a$) planar, ($b$) cylindrical and ($c$) super-cylindrical).}
    \label{fig:intro}
\end{figure}

Fig.~\ref{fig:intro} shows an interesting experiment: a wave from a signal source (on the right) hits three different surfaces made of dipoles spaced evenly apart. When the wave hits the flat surface (sub-figure ($a$)), it acts like a mirror, bouncing the wave back in a tight, focused beam. However, when the surface is curved (sub-figures ($b$) and ($c$)), an insightful behavior can be observed: several beams are spreading out over a larger area. This prompts an exploration into how the design of an \gls{em} surface shape can influence the way it interacts with incoming waves, giving rise to the burgeoning field of conformal or \gls{3d} \gls{ris},\footnote{We use the terms \gls{3d} \gls{ris} and conformal \gls{ris} interchangeably throughout the paper.} sculpted to the unique topography of their hosting environments~\cite{Lin2022}. 

Notwithstanding the progress in advanced printing techniques that facilitate the fabrication of irregular electronic boards, the precision placement and orientation of unit cells on \gls{3d} surfaces present formidable technical challenges~\cite{Budhu22}. These include not only the physical constraints of inter-element spacing, alignment, and orientation in \gls{3d}, but also the intricate mathematical modeling necessary for effective dynamic control of such systems~\cite{Mizmizi23, Zelaya21}. The fidelity of models based on infinite planar arrays falters when confronted with the finite size and curvature of real-world structures, leading to discrepancies between theoretical predictions and empirical observations. Therefore,
the optimization of the \gls{3d} \gls{ris} layout and its corresponding configuration demands sophisticated algorithms tuned to complex environmental dynamics and \gls{em} interactions.

Recently, impedance-based models have been proposed to advance the optimization and efficient design of \glspl{ris} \cite{Gradoni2021,MarcoSergei2022}. It has been observed that spacing the \gls{ris} elements at inter-distances smaller than $\lambda/2$ results in a more compact layout, while still allowing for effective beamforming gains despite the additional mutual coupling among the \gls{ris} elements, i.e., interactions between the elements of the antenna array. When dealing with non-planar \glspl{ris}, exploiting the mutual coupling at the design stage may benefit the \gls{ris} configuration, leading to significant performance gains beyond \gls{sota} solutions~\cite{Nolan21, Ma23}. 

Through a comprehensive investigation of the fundamental properties and performance characteristics of conformal \glspl{ris}, we demonstrate that these \gls{3d} structures can be optimized in their design phase and offer unprecedented capabilities for manipulating \gls{em} waves.
In particular, we propose a novel solution, namely \name{} (Tailored \gls{3d} \gls{ris}, pronounced ``tedris''), wherein the arrangement of \gls{ris} elements is optimized while leveraging the mutual coupling to maximize performance and keeping the \gls{ris} conformal to the support surface. To the best of our knowledge, \name{} is the first attempt at optimizing the layout of \gls{3d} \gls{ris}. Hence, it constitutes a breakthrough in the design of intelligent surfaces, holding great promise for revolutionizing wireless communication systems and advancing several applications for \gls{6g}.

\subsection{Contributions}


The paper's contributions can be elaborated upon as follows:

{\bf C1. Demonstration of 3D RIS Feasibility for Communication:} 
The paper first and foremost establishes the practicality of designing 3D \gls{ris} specifically for communication purposes. This is a crucial step forward as it moves beyond the traditional flat-panel designs, exploring the potential of \gls{ris} in three-dimensional forms. This advancement is critical because \gls{3d} \gls{ris} can potentially conform to various surfaces and shapes, offering more flexibility and adaptability in various environments compared to their 2D counterparts.

{\bf C2. Mutual Coupling-Aware Model for \gls{3d} Conformal \gls{ris}:}
We delve into the complex phenomenon of mutual coupling in \gls{3d} \gls{ris} with structures going beyond planar geometry, a factor often overlooked in previous studies, which can significantly impact performance. By developing a model that accounts for these interactions, especially in \glspl{ris} with non-uniform spacing between elements, our paper provides a more accurate and realistic approach to future \gls{ris} design. Moreover, it introduces significant research opportunities in accurately modeling arbitrary impedance boundary conditions in conformal structures.

{\bf C3. Framework for Joint Optimization of \gls{3d} Shape and Configuration:} 
Building on the mutual coupling-aware \gls{ris} channel model~\cite{Gradoni2021}, this paper introduces a novel framework for jointly optimizing both the physical shape of the \gls{ris} (\gls{3d} shape) and its operational configuration. This holistic approach ensures that the design of the \gls{ris} is not only physically feasible but also operationally effective, taking into account factors like signal strength, direction, and interference. Apart from its immediate relevance to \gls{ris}-assisted wireless networks, the introduced framework, which focuses on shaping an \gls{ris} by determining the precise positioning of its elements, is directly applicable to the development of three other cutting-edge technologies: smart skins~\cite{SmartSkins}, fluid antennas~\cite{FluidAntennas}, and movable antenna systems~\cite{MovableAntennas}. Each of these technologies fundamentally relies on optimizing the positions of the radiating components in antenna arrays. Consequently, this framework is expected to play a significant role in shaping the design of future wireless systems.

{\bf C4. Validation through Simulation and Comparative Analysis:} 
Lastly, we validate our proposed approach using a commercial full-wave \gls{em} simulator, which is crucial for testing the theoretical models in practical scenarios. Additionally, we conduct an extensive numerical simulation study, comparing our method against the most relevant solutions in the field. This comprehensive validation process not only demonstrates the effectiveness of our approach, but also highlights its significance and potential impact in the context of existing solutions.

In summary, this paper represents a substantial step forward in the development of conformal \gls{ris} for wireless applications. Its comprehensive approach, from theoretical modeling to practical validation, addresses key challenges in the field and opens up new possibilities for the design and implementation of advanced smart surface technologies.

{\bf Notation.} We denote matrices and vectors in bold text, while each of their element is indicated in Roman font with a subscript. $(\cdot)^{\tran}$, $(\cdot)^{\herm}$, and $(\cdot)^{*}$ stand for vector or matrix transposition, Hermitian transposition, and the complex conjugate operator, respectively. $\| \cdot \|$ denotes the L$2$-norm in the case of a vector and the spectral norm in the case of a matrix, while the Frobenius norm is denoted by $\| \cdot \|_{\mathrm{F}}$. The operators $\tr(\cdot)$, $\mathrm{diag}(\cdot)$, and $\ceil*{\cdot}$ indicate the trace of a matrix, a diagonal matrix, and the ceiling function, respectively. The gradient operator with respect to the $N\times1$ vector $\vx$ is $\nabla_{\vx}= [\frac{\partial}{\partial x_1}, \dots, \frac{\partial}{\partial x_N}]^{\tran}$. $\Real$ and $\Compl$ denote the sets of real and complex numbers, respectively. Lastly, $\cdot$ and $\times$ represent dot and cross product, respectively.

{\bf Outline of the Paper.}
Section~\ref{sec:physics} gives a comprehensive analysis of the \gls{em} implications arising from the conformal structure of \glspl{ris}, together with their effect on signal propagation. Section~\ref{sec:system_model} details the system model. The formulation of the joint shape and configuration optimization problem for \gls{3d} \gls{ris} is outlined in Section~\ref{sec:problem}, together with our proposed solution, \name{}. The performance evaluation is provided in Section~\ref{sec:results}, followed by an exploration of related works in Section~\ref{sec:related}. Finally, Section~\ref{sec:conclusion} closes the paper with concluding remarks summarizing key insights and contributions.

\section{Fundamentals of Electromagnetic Wave Manipulation in 3D RIS}\label{sec:physics}
We outline four critical questions that address the issue of \gls{3d} \gls{ris} design from an \gls{em} standpoint, which are detailed in the following.

\emph{Q1. How do curved \change{\gls{ris}} affect signal propagation?}

\noindent The answer to this preliminary question has already been provided by~\cite{mattiello2015analysis}: The finite size and non-zero curvature negatively affect the predictions that can be made based on infinite planar array models. Also, there exists a disagreement with measurements, based on spatial filters, such as \gls{fss} structures. 
Specifically, the limitations in modeling accuracy are exacerbated in the case of \glspl{ris} due to the presence of additional re-radiation fields that depend on the non-uniform reflective state of the unit cells. Floquet's theory does not apply to wave propagation in curved surfaces, as the periodicity is broken in the geometrical arrangement of the elements, even if they are equispaced pairwise. 
Hence, full-wave simulation methods based on unit cells terminated by periodic boundary conditions are therefore ruled out.

The only viable option is therefore the mathematical modeling of the whole surface. A comprehensive analysis of modeling methodologies for non-periodic arrays~\cite{mattiello2015analysis} suggests that circuital methods, e.g., based on the \gls{mom} and equivalent \gls{lc} circuits (resonant circuits), constitute a prominent solution in reducing the computation time while providing accurate performance predictions. 
In the following sections, we leverage a discrete model for \glspl{ris} based on the \gls{cma}~\cite{Gradoni2021}. Specifically, transmit and receive antenna arrays, as well as the \glspl{ris} are modeled by leveraging the \gls{dda}~\cite{chaumet2022discrete}, and dipole-to-dipole interactions (the \gls{em} mutual coupling) are modeled by impedance matrices.

%
\emph{Q2. Does mutual coupling play a role in the transition from planar to curved radiating surfaces?}

\noindent In the context of non-periodic surfaces, the mutual coupling between dipoles plays a major role in establishing optimized wave transformations at high efficiency. 
The \gls{mom} provides a way to extend the impedance model beyond \gls{dda} as it treats arbitrary surface current density and devises the radiated field accounting for the influence of nearby unit cells, thus capturing mutual coupling~\cite{gibson2007method}. 
On the one hand, increasing the ability to control the wavefronts requires the design of \glspl{ris} with a large number of unit cells per wavelength~\cite{MarcoSergei2022}. 
On the other hand, non-planar arrays are non-periodic and have a finite size, whence they are also configured as geometrically non-uniform~\cite{mattiello2015analysis}.

Furthermore, \gls{ris} structures introduce non-uniformities on the local reflection coefficient. Thus, the design of \gls{3d} \glspl{ris}
requires novel \gls{em} modeling methods due to their non-uniform and non-planar structure. 
In general, further degrees of freedom in wave control can be gained by intentionally introducing 
non-uniform, and even disordered, arrangements of elements already within the curved shell of a \gls{ris}.
A notable example is constituted by hyper-uniform disordered antenna arrays~\cite{zhang2021hyperuniform}, where the side lobes can be effectively suppressed thanks to the presence of non-uniformity.
However, spurious lobes can be generated by curved surfaces due to \emph{creepy waves}, which are those surface waves that are converted into free-space radiation by propagating along the surface curvature, and for which judicious physical modeling and predictive design are imperative. 
Non-uniform element arrangements are hard to model, but the mutual impedance model introduced in~\cite{Gradoni2021} for planar \gls{ris} structures can inherently tackle non-uniformity. 

\emph{Q3. How does cell and surface curvature affect the radiated wave fields?}

\noindent Suppose a unit cell has a curvature in the order of a wavelength, the impinging field induces an equivalent surface current density whose flow would be perturbed by the non-planar structure. 
This is captured by an impedance boundary condition that can be expressed in the appropriate local frame of reference. 
For the sake of simplicity, let us consider a cylindrical structure with coordinates $(\rho,\phi,z)$ representing the curvature radius, azimuth angle, and elevation, respectively.\footnote{Note that the provided analysis can be generalized for any surface geometry.} Therefore, we can write the surface impedance boundary condition as the following
\begin{equation}
    \textbf{E}(\rho,\phi,z) = \overline{\overline{\mZ}}_s(\phi,z) \cdot \left [\hat{\rhob} \times \Delta \textbf{H}(\rho,\phi,z) \right ],
\end{equation}
which relates the total (boundary) electric field $\textbf{E}(\rho,\phi,z)$ to the magnetic fields at the left $\textbf{H}^{\textrm{l}}(\rho,\phi,z)$ and at the right $\textbf{H}^{\textrm{r}}(\rho,\phi,z)$ of the equivalent impedance sheet modelling the meta-surface, whence  
\begin{equation}
    \Delta \textbf{H}(\rho,\phi,z)= \textbf{H}^{\textrm{l}}(\rho,\phi,z) - \textbf{H}^{\textrm{r}}(\rho,\phi,z),
\end{equation}
and where $\overline{\overline{\mZ}}_s$ is the non-uniform (homogeneous) surface impedance that achieves a prescribed scattering profile~\cite{yang2019surface} and $\hat{\rhob}$ radial versor perpendicular to the surface.
The local scattered field is thus related to the induced current through $\textbf{J}_s(\rho,\phi,z) = \hat{\rhob} \, \times \, \Delta \textbf{H}(\rho,\phi,z)$. 
Using this and decomposing the total field into incident and scattered components, i.e., $\textbf{E}(\rho,\phi,z)=\textbf{E}^{\textrm{inc}}(\rho,\phi,z)+\textbf{E}^{\textrm{sca}}(\rho,\phi,z)$, yields 
\begin{equation}\label{eqn:IBC}
    \textbf{E}^{\textrm{sca}}(\rho,\phi,z) = - \textbf{E}^{\textrm{inc}}(\rho,\phi,z) + \overline{\overline{\mZ}}_s(\phi,z) \cdot \textbf{J}_s(\rho,\phi,z).
\end{equation}
Consequently, the \gls{mom} can be used to devise the radiated fields of curved surfaces through the reaction integrals~\cite[Eqs. (4) and (5)]{Bosiljevac2020}, which are used to devise the elements of the finite size impedance matrix of the meshed surface current~\cite{gibson2007method} as follows 
\begin{equation}
    \textbf{V}^{\textrm{MS}} = \textbf{Z}^{\textrm{MS}} \, \textbf{I}^{\textrm{MS}},
\end{equation}
where the surface port voltages $\textbf{V}^{\textrm{MS}}$ and currents $\textbf{I}^{\textrm{MS}}$ are related by the impedance matrix $\textbf{Z}^{\textrm{MS}}$ that is directly related to the (circuital) active impedance matrix.
Upon expansion of the surface current density onto the basis functions $\textbf{u}_n^{\textrm{MS}}$ defined across a spatial mesh   
\begin{equation}
    \textbf{J}_s = \sum_{n} I^{\textrm{MS}}_{n} \textbf{u}_n^{\textrm{MS}},
\end{equation}
where $I^{\textrm{MS}}_{n}$ is an element of $\textbf{I}^{\textrm{MS}}$ that represents the associated current phasor, and the entries of $\textbf{Z}^{\textrm{MS}}$ configure as a two-part impedance
\begin{equation}\label{eqn:ZMS}
    Z_{ij}^{\textrm{MS}}(n,m) = Z_{ij,G}^{\textrm{MS}} (n,m) + Z_{ij,Z_s}^{\textrm{MS}} (n,m),
\end{equation}
with individual parts given by an overlapping integral across the surface space, given as
\begin{equation}
    Z_{ij,G}^{\textrm{MS}} (n,m) = \eta \, \int u_{i}^{\textrm{MS}} (n) \, G_{ij} (n,m) \, u_{j}^{\textrm{MS}} (m) \,\, dS,
\end{equation}
where $G_{ij} (n,m)$ is the component $ij$ of the free-space dyadic Green function propagating the fields from current element $n$ to current element $m$, and with
\begin{equation}
    Z_{ij,Z_s}^{\textrm{MS}} (n,m) = \eta \, \int u_{i}^{\textrm{MS}} (n) \, \overline{\overline{Z}}_{s,ij} (n,m) \, u_{j}^{\textrm{MS}} (m) \,\, dS,
\end{equation}
with free-space wave impedance $\eta = 377 \, \Omega$, and where the surface impedance boundary condition couples current element $n$ with current element $m$ through the component $ij$ of the impedance tensor in~\eqref{eqn:IBC}. 
A detailed example for a circular modulated surface has been devised in~\cite{Craeye2019} by using entire-domain basis functions.
In reflect-array and meta-surfaces, this current follows the geometrical shape of the unit cell and is influenced by the dielectric substrate, ground plane (in case the surface is reflective), and embedded circuital element (if we prescribe unit cell reconfiguration). 
This is captured by a modified expression of Green's function that includes multiple reflections between the unit cell interface and the ground plane through the (lossy) dielectric substrate. 
While this is a computationally intensive task, a fast algorithm for general - including curved - multilayered media came to fruition over the last year~\cite{konno2016fast}.
Again, the \gls{mom} provides the most general model to calculate the finite size impedance matrix from arbitrary radiation structures.
The case of planar \gls{ris} has been covered in~\cite{konno2023generalised}, where it is found that the \gls{e2e} impedance model holds also in the case of real-life surface planar structures. 
As in~\cite{konno2023generalised}, also in the case of curved surfaces, the basis functions can be related to a passive element, thus representing a mesh element of the surface boundary impedance, or a port element, which represents the terminals loaded by the (\gls{ris}) control circuitry. 
Therefore, in Section~\ref{sec:system_model} we introduce the \gls{ris} self impedance $\mZ_{\rmS\rmS}$ described 
as the port impedance $\mZ_{ij}^{\textrm{MS}}$ in~\eqref{eqn:ZMS}.

\emph{Q4. How can the impedance model be valid for real-life surface prototypes?}

\noindent It is worth pointing out that the \gls{e2e} impedance model of \gls{ris}-aided wireless channels can be used for arbitrary real-life structures~\cite{mursia2023empirical, zheng2024mutual}. 
Besides the ability of the \gls{mom} to tackle arbitrary \gls{ris} unit cells beyond the \gls{dda} described above, there are further practical aspects that are captured by impedance-based channel models just by extending the \gls{mom} formulation. 
For instance, the presence of a supporting structure (walls, ground plane, etc.) can be accounted for by using the multi-layer Green's functions~\cite{konno2016fast} in the reaction integrals of the \gls{mom} in~\cite{konno2023generalised}. 
Inherently, the surface current densities supported by arbitrary radiating structures usually entail characteristic modes. 
In the case of \gls{ris} structures, two families of modes are included in characteristic modes: i) structural modes, describing the radiation from the base composite structure; ii) re-radiation modes, capturing the unit cell reconfiguration and controlled by the tuning circuitry.
We expect that the radiation from curved \gls{ris} structures includes also a contribution from creepy waves.
We stress that the \gls{mom} predicts the full-wave radiation from arbitrarily curved surfaces, but it remains a challenge to understand how to control or even suppress the spurious radiation from creepy waves. 

Hereafter, we assume that the \gls{ris} elements are electrically small and the radius of curvature \emph{does not bend the geometry of the unit cell} and hence does not change its resonance frequency.
This is an important aspect for the high radius of curvature of planar structure with patch-like unit cells, as pointed out in~\cite{Vukovic2022}, and will be investigated in future developments of the present work. 
Furthermore, in order to make the treatment general and accessible to communication-oriented studies, no substrate is assumed and a dense arrangement of freestanding dipoles is considered, including planar simultaneous transmitting and receiving \glspl{ris} and omnisurfaces that have been recently proposed in the literature to control the \gls{3d} space~\cite{STAR-RIS, OMNI-RIS}. 
Interestingly, the local planar structure approximation does not hold, as there might not be a convenient system of reference where the array is uniform. 
Consequently, approaches based on the vector-Legendre transformation may become invalid~\cite{sipus2007analysis}. 
Based on such considerations, in the next sections, we introduce a new model, formulate an optimization problem, and describe a corresponding optimization algorithm to design mutually coupled non-planar \glspl{ris}.

\section{System model}\label{sec:system_model}

We consider an \gls{ris}-aided network, wherein a single-antenna \gls{bs} communicates with a single-antenna \gls{ue} via an $N$-element \gls{ris}. The center-point of the \gls{ris} is located at the origin, while the \gls{bs} and the \gls{ue} are located in positions $\vp_{\scriptscriptstyle \mathrm{BS}}\in\Real^{3\times 1}$ and $\vp_{\scriptscriptstyle \mathrm{UE}}\in\Real^{3\times 1}$ with respect to the origin (the center-point of the \gls{ris}), respectively. The selected scenario allows us to better focus on the \gls{3d} \gls{ris} layout optimization problem, without losing generality. More complex scenarios with multiple-antenna \gls{bs} and multiple \glspl{ue} under mobility conditions are left for future work. We adopt the \gls{em}-consistent and mutual coupling aware channel model in~\cite{Gradoni2021}, according to which each \gls{ris} element is modeled as a thin wire dipole driven by a tunable impedance. Specifically, the end-to-end channel is given by
\begin{align}\label{eq:H}
    H = Y_0[Z_{\rmR\rmT} - \vz_{\rmS\rmR}^{\tran}(\mZ_{\rmS\rmS}+\mZ_{\rmRIS})^{-1}\vz_{\rmS\rmT}] \in \Compl
\end{align}
where $Y_0 = Z_{\rmL}(Z_{\rmL}+Z_{\rmR\rmR})^{-1}(Z_{\rmG}+Z_{\rmT\rmT})^{-1} \in \Compl$ is a fixed constant, 
the subscripts $\mathrm{R},~\mathrm{T},~\mathrm{S}$ stand for the \gls{ue}, the \gls{bs} and the \gls{ris}, respectively, and $Z_L$ and $Z_G$ are the load impedances of the \gls{ue} and the \gls{bs}, respectively. Moreover, $Z_{\rmR\rmR}$ and $Z_{\rmT\rmT}$ denote the self impedances at the \gls{ue} and \gls{bs}, which are assumed to be known. $Z_{\rmR\rmT}\in\Compl$ represents the direct channel between the \gls{bs} and the \gls{ue}, 
$\vz_{\rmS\rmR}\in\Compl^{N\times 1}$
stands for the channel between the \gls{ris} and the \gls{ue}, and $\vz_{\rmS\rmT}\in\Compl^{N\times 1}$ denotes the channel between the \gls{bs} and the \gls{ris}. Lastly, $\mZ_{\rmS\rmS}\in\Compl^{N\times N}$ denotes the matrix of self and mutual impedances among the elements of the \gls{ris} (derived as the port impedance  $\mZ_{ij}^{\textrm{MS}}$ in~\eqref{eqn:ZMS}), and $\mZ_{\rmRIS}\in \Compl^{N\times N}$ is the matrix of tunable \gls{ris} impedances. 

The received signal at the \gls{ue} is thus given by
\begin{align}
    y = H x + n \in \Compl
\end{align}
where $x$ is the transmitted symbol, $\Exp[|x|^2] = P$ with $P$ being the total transmit power, and $n$ is the noise at the receiver that is distributed as $\mathcal{CN}(0,\sigma_n^2)$. 

The \gls{ris} tunable impedance matrix is modeled as
\begin{align}
    \mZ_{\rmRIS} = \mathrm{diag}\{R_0+j b_n\}_{n=1}^N \label{eq:Zris}
\end{align}
where $R_0\geq 0$ is a constant that depends on the losses of the \gls{ris} internal circuits, and $b_n\in \Real$ is the tunable reactance at the $n$-th RIS element, i.e., the configuration of the $n$-th element.

The entries of the matrix $\mZ_{\rmS\rmS}\in\Compl^{N\times N}$ can be computed by utilizing the analytical framework introduced in~\cite{DiRenzo23}.\footnote{Full-wave simulations can be used to compute the $Z$ parameters required by our optimization approach, with no assumption on the element type or scattering conditions~\cite{Bosiljevac2020}. Nonetheless, dipoles and \gls{cma} allow us to approach the problem analytically~\cite{Gradoni2021}.} Likewise, the same framework can be utilized to compute the vectors $\vz_{\rmR\rmS}$ and $\vz_{\rmS\rmT}$, as well as $Z_{\rmR\rmT}$. Specifically, the self and mutual impedance between any pair of thin wire dipoles $q$ and $p$ of length $\lambda/2$ and with position $\vq_q\in\Real^3$ and $\vq_p\in\Real^3$, respectively, can be formulated as (see Appendix~\ref{ap:A1} for the details)
\begin{align} \label{eq:zqp_compact}
Z_{qp} = \frac{\eta}{8\pi} &\sum_{s_o = \{-1, 1 \}} e^{j \frac{2\pi}{\lambda} s_0 \ve_3^{\tran}\deltab_{qp}}\big[T_0(\deltab_{qp}-2h\ve_3, s_0)+\nonumber\\
&T_0(\deltab_{qp}+2h\ve_3, s_0)-2T_0(\deltab_{qp}, s_0)\big],
\end{align}
where $\eta=377~\Omega$ is the free-space wave impedance, $\lambda$ is the signal wavelength, $h$ is the half-length of the dipoles, and $\deltab_{qp} = \vq_q - \vq_p$.
Moreover, we have defined the function
\begin{align}
T_0(\zetab,s_0) = E_1 \Big(jk\big(\|\zetab \|+s_0\ve_3^{\tran} \zetab\big)\Big),\label{eq:T}
\end{align}
where $s_0\in\{-1,1\}$, and $E_1(z)$ is the exponential integral defined as
\change{$E_1(z) = \int_{z}^\infty \frac{e^{-t}}{t} dt$.}

The expression in \eqref{eq:zqp_compact} is not applicable when the dipoles $q$ and $p$ are co-linear~\cite{DiRenzo23}. In that case, the expression in~\cite[Eq. (9)]{DiRenzo23} can be utilized via numerical integration.

\section{Problem formulation}\label{sec:problem}

We aim at maximizing the \gls{snr} at the \gls{ue} by finding the optimal values of the tunable impedances driving the \gls{ris} elements and the optimal locations of the \gls{ris} elements, i.e., the \gls{ris} \emph{configuration} and the \gls{ris} \emph{shape}, respectively. To this end, we denote the vector of \gls{ris} configurations as $\vb = [b_1,\ldots,b_N]^{\tran}\in \Real^{N\times 1}$ and the set of \gls{ris} positions as $\mQ = [\vq_1,\ldots,\vq_N] \in \Real^{3\times N}$, which represent the optimization variables, accordingly. Note that in~\eqref{eq:H}, $\vz_{\rmS\rmR}$ and $\vz_{\rmS\rmT}$ depend on $\mQ$, and they can be computed by using~\eqref{eq:zqp_compact}.

Therefore, we focus on finding the optimal $\vb$ and $\mQ$ by solving the following optimization problem
\begin{problem}\label{eq:P1}
\begin{align}
         \displaystyle \max_{\vb,\,\mQ} \,\, & |Z_{\rmR\rmT} \!-\! \vz_{\rmS\rmR}^{\tran}(\mQ)(\mZ_{\rmS\rmS}(\mQ)\!+\!\mZ_{\rmRIS}(\vb))^{-1}\vz_{\rmS\rmT}(\mQ)|^2  \\
         & \mathrm{s.t.} \quad \mQ \in \mathcal{Q}^N,\quad \vb\in\mathcal{B}^N, 
\end{align}    
\end{problem}
\noindent where $\mathcal{Q}$ and $\mathcal{B}$ represent the feasible sets of the \gls{ris} elements positions and tunable reactances, respectively. Note also that we assume $\mathcal{Q}$ to be a convex set (see Sections~\ref{subsec:3Dsphere}--\ref{subsec:cyl}). It is worth specifying that the selection of the feasible set $\mathcal{Q}$ is dictated by the deployment constraints of the surface. In other words, it ensures the practicality of the obtained solution. However, constrained sets might lead to suboptimal element arrangements from a pure communication standpoint. Nevertheless, the selection of an unconstrained set yields an optimal \gls{ris} layout for communication, which can be a valuable input for defining new strategies for the design of conformal \glspl{ris}.

Problem~\ref{eq:P1} is particularly challenging to solve given its non-convex formulation, the inversion of the matrix $\mZ_{\rmS\rmS}(\mQ)+\mZ_{\rmRIS}(\vb)$, and the fact that, except for $Z_{\rmR\rmT}$, the impedances to be optimized depend on the locations of the \gls{ris} elements $\mQ$. Moreover, we emphasize that \gls{sota} research works focused only on the optimization of the \gls{ris} tunable reactances~\cite{Mursia23,Hassani23,Qian20,Abrardo21}, while no research work has tackled the joint optimization of the \gls{ris} configuration and shape.

\subsection{\name{} Algorithm}

Hereafter, we introduce the proposed iterative algorithm to tackle Problem~\ref{eq:P1}, which is referred to as \emph{\name{}}.

Let us denote the position of the \gls{ris} elements at the $i$-th iteration of \name{} as $\mQ^{(i)}$, and the corresponding value of the mutual impedances between the transmitter and the \gls{ris}, the self and mutual impedances between the \gls{ris} elements, and between the latter and the receiver as $\vz_{\rmS\rmR}^{(i)} = \vz_{\rmS\rmR}(\mQ^{(i)})$, $\mZ_{\rmS\rmS}^{(i)}=\mZ_{\rmS\rmS}(\mQ^{(i)})$ and $\vz_{\rmS\rmT}^{(i)}=\vz_{\rmS\rmT}(\mQ^{(i)})$. Note that we depart from an initial condition in which the \gls{ris} elements are located at the initial position $\mQ^{(0)}$, and the corresponding self and mutual impedances are given by $\vz_{\rmS\rmR}^{(0)}$, $\mZ_{\rmS\rmS}^{(0)}$ and $\vz_{\rmS\rmT}^{(0)}$.\footnote{The initial shape $\mQ^{(0)}$ can be arbitrarily chosen without affecting the overall performance of the proposed algorithm, as shown in Section~\ref{sec:results}.}

Therefore, we focus on optimizing the \gls{ris} configuration, which is specified by~\eqref{eq:Zris}. Specifically, at each iteration $i$ of the proposed algorithm, we formulate the following optimization problem
\begin{problem}\label{eq:P2}
\begin{align}
         \max_{\mZ_{\rmRIS}} & |Z_{\rmR\rmT} -\vz_{\rmS\rmR}^{\tran(i)}(\mZ_{\rmS\rmS}^{(i)}+\mZ_{\rmRIS})^{-1}\vz_{\rmS\rmT}^{(i)}|^2  \\
         & \mathrm{s.t.}  \mZ_{\rmRIS} \!=\! \mathrm{diag}(R_0 + j b_k), b_k \in \mathcal{B}, \, k \!=\!1,\ldots,N,
\end{align}    
\end{problem}
\noindent which can be solved by using \cite{Mursia23, Hassani23}.

Once the optimized values of $\mZ_{\rmRIS}$ are obtained, we turn to the optimization of the positions of the \gls{ris} elements by formulating the following optimization problem
\begin{problem}\label{eq:P3}
    \begin{align}
         \max_{\mQ}~ & |Z_{\rmR\rmT} -\vz_{\rmS\rmR}^{\tran(i)}(\mZ_{\rmS\rmS}^{(i)}+\mZ_{\rmRIS})^{-1}\vz_{\rmS\rmT}^{(i)}|^2  \nonumber \\
         & \mathrm{s.t.}~  \mQ \in \mathcal{Q}^N,
\end{align}
\end{problem}
which we tackle by using the projected gradient ascent method, as
\begin{align}
    \mQ^{(i+1)} = P_{\mathcal{Q}}\Big(\mQ^{(i)} + \alpha_i \nabla f(\mQ^{(i)})\Big), \label{eq:proj_descent}
\end{align}
where $P_{\mathcal{Q}}$ is the projection operator onto the feasible set $\mathcal{Q}$, $f(\mQ^{(i)})$ is the objective function in Problem~\ref{eq:P3}, which is expressed as a function of $\mQ^{(i)}$, and $\alpha_i>0\in\Real$ is the step size. We define $P_{\mathcal{Q}}(\mQ)$ as 
\begin{align}
    P_{\mathcal{Q}}(\mQ_0) = \displaystyle \arg \min_{\mQ\in\mathcal{Q}^N} \frac{1}{2} \|\mQ-\mQ_0\|^2_{\mathrm{F}}.
\end{align}
Hence, the updating rule corresponds to solving the following problem
\begin{align}
     \displaystyle \min_{\mQ^{(i+1)}\in\mathcal{Q}^N} \|\mQ^{(i+1)}-\mQ^{(i)} - \alpha_i \nabla f(\mQ^{(i)})\|_{\mathrm{F}}^2.
\end{align}

The gradient of $f(\mQ^{(i)})$ is challenging to compute due to the expression of each $Z_{qp}$ in \eqref{eq:zqp_compact} and the fact that each $Z_{qp}$ appears inside the inverse of $\mZ_{\rmS\rmS}$. Hence, we employ the Neumann series approximation to tackle this problem~\cite{Qian20}.
Specifically, at each iteration of the algorithm, we optimize only a small variation of the position of the \gls{ris} elements as $\mQ^{(i+1)} = \mQ^{(i)} + \Delta\mQ$, where $\Delta\mQ\in\Real^{3\times N}$ denotes the small variations.
More precisely, we update the location of one \gls{ris} element at a time. Hence, for each $k=1,\ldots,N$, we have $\vq_k^{(i+1)} = \vq_k^{(i)}+\Delta\vq_k$. As a result, at the $i$-th iteration of the algorithm, we obtain  $\mZ_{\rmS\rmS}^{(i+1)} = \mZ_{\rmS\rmS}^{(i)}+ \Deltab_{\rmS\rmS}^{(k)},~\forall k$, where the matrix $\Deltab_{\rmS\rmS}^{(k)}\in\Compl^{N\times N}$ is non-zero only in $k$-th row and column, which correspond to the $k$-th \gls{ris} element being updated.
Moreover, due to $\mZ_{\rmS\rmS}$ being symmetric by definition, also $\Deltab_{\rmS\rmS}^{(k)} = (\Deltab_{\rmS\rmS}^{(k)})^{{\tran}}$ applies. Similarly, we have
\begin{align}
    {\vz_{\rmS\rmR}^{(i+1)} = \vz_{\rmS\rmR}^{(i)} + \deltab_{\rmS\rmR}^{(k)}} \quad \text{and} \quad
    \vz_{\rmS\rmT}^{(i+1)} = \vz_{\rmS\rmT}^{(i)} + \deltab_{\rmS\rmT}^{(k)},
\end{align}
where only the $k$-th entry of $\deltab_{\rmS\rmR}^{(k)}\in\Compl^{N\times 1}$ and $\deltab_{\rmS\rmT}^{(k)}\in\Compl^{N\times 1}$ is non-zero.

Let $\mG^{(i)} = (\mZ_{\rmS\rmS}^{(i)} + \mZ_{\rmRIS})^{-1}$, and using the Neumann series approximation, the inverse matrix in Problem~\ref{eq:P3} is approximated as
\begin{align}
    (\mZ_{\rmS\rmS}^{(i)} + \mZ_{\rmRIS})^{-1}\approx \mG^{(i)} - \mG^{(i)}\Deltab_{\rmS\rmS}^{(k)}\mG^{(i)} = \change{\Tilde{\Gammab}^{(i)}}, \label{eq:neumann}
\end{align}
provided that $\|\mG^{(i)}\Deltab^{(k)}_{\rmS\rmS}\|\ll 1$~\cite{Qian20,Abrardo21}. Furthermore, note that $\mG^{(i)} = (\mG^{(i)})^{\tran}$.

For ease of writing, we refer to the objective function in Problem~\ref{eq:P3} as
\begin{align}
f(\Delta \vq_k) &= | h(\Delta \vq_k) |^2 \nonumber\\
&=| Z_{\rmR\rmT} - \vz_{\rmS\rmR}^{\tran(i+1)}(\mZ_{\rmS\rmS}^{(i+1)}+\mZ_{\rmRIS})^{-1}\vz_{\rmS\rmT}^{(i+1)}|^2.
\end{align}
From \eqref{eq:neumann}, we can approximate the function $h(\Delta \vq_k)$ as
\begin{align}
    h(\Delta \vq_k) \approx \,& Z_{\rmR\rmT} -  (\vz_{\rmS\rmR}^{(i)} + \deltab_{\rmS\rmR}^{(k)})^{\tran}\change{\Tilde{\Gammab}^{(i)}}(\vz_{\rmS\rmT}^{(i)} + \deltab_{\rmS\rmT}^{(k)}). \label{eq:obj_approx}
\end{align}
Moreover, we introduce the variables
\begin{align}
    Z_{\rmR\rmS\rmT}^{(i)} &  = Z_{\rmR\rmT} -\vz_{\rmS\rmR}^{\tran(i)}\mG^{(i)}\vz_{\rmS\rmT}^{(i)}\in\Compl,\\
    \vg_{\rmS\rmT}^{(i)} &=  \mG^{(i)}\vz_{\rmS\rmT}^{(i)}\in\Compl^{N\times 1},\\
    \vg_{\rmS\rmR}^{(i)} & =\mG^{(i)}\vz_{\rmS\rmR}^{(i)}\in\Compl^{N\times 1},
\end{align}
and rewrite \eqref{eq:obj_approx} in a more convenient form as
\begin{align}
    h&(\Delta \vq_k) \approx Z_{\rmR\rmS\rmT}^{(i)} - (\deltab_{\rmS\rmR}^{(k)})^{\tran} \vg_{\rmS\rmT}^{(i)}+(\vg_{\rmS\rmR}^{(i)})^{\tran}\Deltab_{\rmS\rmS}^{(k)}\vg_{\rmS\rmT}^{(i)}\nonumber \\
    &+(\deltab_{\rmS\rmR}^{(k)})^{\tran}\mG^{(i)}\Deltab_{\rmS\rmS}^{(k)}\vg_{\rmS\rmT}^{(i)}-(\vg_{\rmS\rmR}^{(i)})^{\tran}\deltab_{\rmS\rmT}^{(k)} \nonumber \\
    & -(\deltab_{\rmS\rmR}^{(k)})^{\tran}\mG^{(i)}\deltab_{\rmS\rmT}^{(k)}+ (\vg_{\rmS\rmR}^{(i)})^{\tran}\Deltab_{\rmS\rmS}^{(k)}\mG^{(i)}\deltab_{\rmS\rmT}^{(k)}\nonumber \\
    &+ (\deltab_{\rmS\rmR}^{(k)})^{\tran}\mG^{(i)}\Deltab_{\rmS\rmS}^{(k)}\mG^{(i)}\deltab_{\rmS\rmT}^{(k)}.
\end{align}

Also, the matrix $\Deltab_{\rmS\rmS}^{(k)}$ can be expressed as 
\begin{align}
    \Deltab_{\rmS\rmS}^{(k)} = \deltab_{\rmS\rmS}^{(k)} \ve_k^{\tran} + \ve_k (\deltab_{\rmS\rmS}^{(k)})^{\tran} - \delta_{k,\rmS\rmS}^{(k)}\ve_k\ve_k^{\tran},
\end{align}
where $\deltab_{\rmS\rmS}^{(k)}\in\Compl^{N\times 1}$ is the only non-zero column of $\Deltab_{\rmS\rmS}^{(k)}$, $\delta_{k,\rmS\rmS}^{(k)}\in\Compl$ is the element in position $(k,k)$ of $\Deltab_{\rmS\rmS}^{(k)}$, or equivalently the only non-zero element of $\deltab_{\rmS\rmS}^{(k)}$, and $\ve_k$ is the $k$-th column of the identity matrix of size $N$. Hence, we obtain
\begin{align}
{(\vg_{\rmS\rmR}^{(i)})^{\tran}}\Deltab_{\rmS\rmS}^{(k)}\vg_{\rmS\rmT}^{(i)}&  = {(\vg_{\rmR\rmS\rmT}^{(i)})^{\tran}}\deltab_{\rmS\rmS}^{(k)}\\
    (\deltab_{\rmS\rmR}^{(k)})^{\tran}\mG^{(i)}\Deltab_{\rmS\rmS}^{(k)}\mG^{(i)}\deltab_{\rmS\rmT}^{(k)}& = \delta_{k,\rmS\rmR}^{(k)}(\vg_k^{(i)})^{\tran}\Deltab_{\rmS\rmS}^{(k)}\vg_k^{(i)}\delta_{k,\rmS\rmT}^{(k)} \nonumber \\&=\delta_{k,\rmS\rmR}^{(k)}(\bar{\vg}_{k}^{(i)})^{\tran}\deltab_{\rmS\rmS}^{(k)}\delta_{k,\rmS\rmT}^{(k)},
\end{align}
where $\vg_{k}^{(i)}$ is the $k$-th column of $\mG^{(i)}$, and we have defined {$\vg_{\rmR\rmS\rmT}^{(i)} = g_{k,\rmS\rmT}^{(i)}\vg_{\rmS\rmR}^{(i)}+g_{k,\rmS\rmR}^{(i)}\vg_{\rmS\rmT}^{(i)}-g_{k,\rmS\rmT}^{(i)}g_{k,\rmS\rmR}^{(i)}\ve_k \in \Compl^{N\times1}$} and $\bar{\vg}_{k}^{(i)} = g_{k,k}^{(i)}\vg_{k}^{(i)}+g_{k,k}^{(i)}\vg_{k}^{(i)}-g_{k,k}^{(i)}g_{k,k}^{(i)}\ve_k$.
Similarly, we obtain
\begin{align}
    \mG^{(i)}\Deltab_{\rmS\rmS}^{(k)}\vg_{\rmS\rmT}^{(i)} & = {(\widetilde{\mG}_{\rmS\rmT}^{(i)})^{\tran}}\deltab_{\rmS\rmS}^{(k)},\\
    {(\vg_{\rmS\rmR}^{(i)})^{\tran} \Deltab_{\rmS\rmS}^{(k)}\mG^{(i)}} & = {(\deltab_{\rmS\rmS}^{(k)})^{\tran}\widetilde{\mG}_{\rmS\rmR}^{(i)} },
\end{align}
where the matrices $\widetilde{\mG}_{\rmS\rmT}^{(i)}$ and $\widetilde{\mG}_{\rmR\rmS}^{(i)}$ are constructed as
\begin{align}
    \widetilde{\mG}_{\rmS\rmT}^{(i)} & = \begin{bmatrix}
    g_{1,\rmS\rmT}^{(i)} {(\vg_{1}^{(i)})^{\tran}} +g_{11}^{(i)}(\vg_{\rmS\rmT}^{(i)})^{\tran}-g_{1,\rmS\rmT}^{(i)}g_{11}^{(i)}\ve_k^{\tran}\\
    \vdots \\
    g_{N,\rmS\rmT}^{(i)}{(\vg_{N}^{(i)})^{\tran}}+g_{NN}^{(i)}(\vg_{\rmS\rmT}^{(i)})^{\tran}-g_{N,\rmS\rmT}^{(i)}g_{NN}^{(i)}\ve_k^{\tran}
    \end{bmatrix}^{\tran} \nonumber \\&= \begin{bmatrix}
    \tilde{\vg}_{1,\rmS\rmT}^{(i)}&
    \hdots &
    \tilde{\vg}_{N,\rmS\rmT}^{(i)}
    \end{bmatrix}\in\Compl^{N\times N},\\
    \widetilde{\mG}_{\rmS\rmR}^{(i)}& = \begin{bmatrix}
    g_{1,\rmS\rmR}^{(i)}{(\vg_{1}^{(i)})^{\tran}}+g_{11}^{(i)}{(\vg_{\rmS\rmR}^{(i)})^{\tran}}-g_{1,\rmS\rmR}^{(i)}g_{11}^{(i)}\ve_k^{\tran}\\
    \vdots \\
    g_{N,\rmS\rmR}^{(i)}{(\vg_{N}^{(i)})^{\tran}}+g_{NN}^{(i)}{(\vg_{\rmS\rmR}^{(i)})^{\tran}}-g_{N,\rmS\rmR}^{(i)}g_{NN}^{(i)}\ve_k^{\tran}
    \end{bmatrix}^{\tran}\nonumber\\&=\begin{bmatrix}
    \tilde{\vg}_{1,\rmS\rmR}^{(i)}&
    \hdots &
    \tilde{\vg}_{N,\rmS\rmR}^{(i)}
    \end{bmatrix}\in\Compl^{N\times N}.
\end{align}

Since the locations of the \gls{ris} elements are updated one at a time, we can further simplify the approximate objective function of Problem~\ref{eq:P3} in \eqref{eq:obj_approx} by expressing it as a function of the columns of $\widetilde{\mG}_{\rmS\rmT}^{(i)}$ and $\widetilde{\mG}_{\rmS\rmR}^{(i)}$ as
\begin{align}
    h&(\Delta \vq_k) \approx  Z_{\rmR\rmS\rmT}^{(i)}- \delta_{k,\rmS\rmR}^{(k)} g_{k,\rmS\rmT}^{(i)}+(\vg_{\rmR\rmS\rmT}^{(i)})^{\tran}\deltab_{\rmS\rmS}^{(k)}\nonumber \\
    &+\delta_{k,\rmS\rmR}^{(k)}(\tilde{\vg}_{k,\rmS\rmT}^{(i)})^{\tran}\deltab_{\rmS\rmS}^{(k)} -g_{k,\rmS\rmR}^{(i)}\delta_{k,\rmS\rmT}^{(k)} -G_{k,k}^{(i)}\delta_{k,\rmS\rmR}^{(k)}\delta_{k,\rmS\rmT}^{(k)}\nonumber \\
    &+ (\deltab_{\rmS\rmS}^{(k)})^{\tran}\tilde{\vg}_{k,\rmS\rmR}^{(i)}\delta_{k,\rmS\rmT}^{(k)}+ \delta_{k,\rmS\rmR}^{(k)}(\bar{\vg}_{k}^{(i)})^{\tran}\deltab_{\rmS\rmS}^{(k)}\delta_{k,\rmS\rmT}^{(k)}. \label{eq:h_max_reduction}
\end{align}
Hence, after applying the Neumann series we can rewrite Problem~\ref{eq:P3} as the following
\begin{problem}\label{eq:P4}
    \begin{align}
         \max_{\Delta \vq_k}~ & \big\rvert Z_{\rmR\rmS\rmT}^{(i)}- \delta_{k,\rmS\rmR}^{(k)} g_{k,\rmS\rmT}^{(i)}+(\vg_{\rmR\rmS\rmT}^{(i)})^{\tran}\deltab_{\rmS\rmS}^{(k)}+\delta_{k,\rmS\rmR}^{(k)}(\tilde{\vg}_{k,\rmS\rmT}^{(i)})^{\tran}\deltab_{\rmS\rmS}^{(k)}\nonumber \\
         & -g_{k,\rmS\rmR}^{(i)}\delta_{k,\rmS\rmT}^{(k)} -G_{k,k}^{(i)}\delta_{k,\rmS\rmR}^{(k)}\delta_{k,\rmS\rmT}^{(k)}\nonumber \\
         &+ (\deltab_{\rmS\rmS}^{(k)})^{\tran}\tilde{\vg}_{k,\rmS\rmR}^{(i)}\delta_{k,\rmS\rmT}^{(k)}+ \delta_{k,\rmS\rmR}^{(k)}(\bar{\vg}_{k}^{(i)})^{\tran}\deltab_{\rmS\rmS}^{(k)}\delta_{k,\rmS\rmT}^{(k)}\big\rvert^2, \,\forall k\nonumber \\
         & \mathrm{s.t.}\quad  \Delta\vq_k\, : \, \|\mG^{(i)}\Deltab^{(k)}_{\rmS\rmS}\|\ll 1, 
\end{align} \end{problem}
where the constraint is needed to make the Neumann series approximation accurate by design.
We are now in the position of computing the gradient of the approximate objective function of Problem \ref{eq:P3} in \eqref{eq:proj_descent} with respect to the (small) update on the position of the $k$-th \gls{ris} element.
We remark that $\vq_k\in\Real^{3\times 1}$ is the $k$-th column of the matrix $\mQ\in\Real^{3\times N}$, which contains the position of the \gls{ris} elements and $\Delta\vq_k$ represents its update.

The gradient of $f(\Delta \vq_k)$ is given by the following
\begin{align}
    \nabla f&(\Delta\vq_k) = \nabla |h(\Delta\vq_k)|^2 = \nabla \big(h(\Delta\vq_k)h^*(\Delta\vq_k)\big) = \nonumber \\
    & =  \big(\nabla h(\Delta\vq_k)\big)h^*(\Delta\vq_k) + h(\Delta\vq_k)\big(\nabla^* h(\Delta\vq_k)\big) .\label{eq:nabla_f}
\end{align}
From the expression in \eqref{eq:h_max_reduction}, the gradient of $h(\Delta\vq_k)$ is approximated by
\begin{align}
&\nabla h(\Delta\vq_k) \approx \nonumber \\
&\nabla\delta_{\rmS\rmR}^{(k)} \bigg[(\tilde{\vg}_{k,\rmS\rmT}^{(i)})^{\tran}\deltab_{\rmS\rmS}^{(k)}\!-\!g_{k,\rmS\rmT}^{(i)}\!-\!G_{k,k}^{(i)}\delta_{k,\rmS\rmT}^{(k)}\!+\!(\bar{\vg}_{k}^{(i)})^{\tran}\deltab_{\rmS\rmS}^{(k)}\delta_{k,\rmS\rmT}^{(k)}\bigg] \nonumber \\
&+  \nabla\delta_{\rmS\rmT}^{(k)} \bigg[(\deltab_{\rmS\rmS}^{(k)})^{\tran}\tilde{\vg}_{k,\rmS\rmR}^{(i)}\!+\!(\bar{\vg}_{k}^{(i)})^{\tran}\deltab_{\rmS\rmS}^{(k)}\delta_{\rmS\rmR}^{(k)}\!-\!g_{k,\rmS\rmR}^{(i)}\!-\!G_{k,k}^{(i)}\delta_{k,\rmS\rmR}^{(k)}\bigg] \nonumber \\
     & +\nabla(\deltab_{\rmS\rmS}^{(k)})^{\tran} \bigg[\vg_{\rmR\rmS\rmT}^{(i)}\!+\!\delta_{k,\rmS\rmR}^{(k)}\tilde{\vg}_{k,\rmS\rmT}^{(i)}\!\!+\!\delta_{k,\rmS\rmR}^{(k)}\delta_{\rmS\rmT}^{(k)}\bar{\vg}_{k}^{(i)}\!\!+\!\tilde{\vg}_{k,\rmS\rmR}^{(i)}\delta_{k,\rmS\rmT}^{(k)}\bigg],\label{eq:partial_h}
\end{align}
while the gradient of $h^*(\Delta\vq_k)$ is obtained from \eqref{eq:partial_h} as follows
\begin{align}\label{eq:partial_hconj}
& \nabla h^*(\Delta\vq_k) = \big(\nabla h(\Delta\vq_k)\big)^*.
\end{align}
In addition, we note that the gradient of $\delta_{\rmS\rmR}^{(k)}$, $\delta_{\rmS\rmT}^{(k)}$ and $\deltab_{\rmS\rmS}^{(k)}$, as well as their complex conjugates, boil down to the gradient of the single impedances $Z_{qp}$. In particular, we have that
\begin{align}
    \nabla\delta_{\rmS\rmR}^{(k)} = \nabla Z_{k \rmR}, \quad \nabla\delta_{\rmS\rmT}^{(k)} = \nabla Z_{k\rmT}, \quad \text{and}
    \nonumber \\ \nabla(\deltab_{\rmS\rmS}^{(k)})^{\tran} = \begin{bmatrix}
      \nabla Z_{1,k}^{(k)},\ldots,\nabla Z_{N,k}^{(k)}\label{eq:partial_delta_Z}
    \end{bmatrix},
\end{align}
where $k$ is the index of the $k$-th \gls{ris} element. Finally, the gradient of $Z_{kl}$ for any two antennas $l$ and $k$, with respect to the antenna $k$ is formulated for $\lambda/2$-length dipoles as
\begin{align}
\nabla Z_{kl} = \frac{\eta}{8\pi} \!\!&\sum_{s_o = \{-1, 1 \}} \!\!\Big( j k s_0 \ve_3 e^{j k s_0 \ve_3^{\tran}(\vq_k - \vq_l))}g(\vq_k - \vq_l, s_0) 
\nonumber \\
&+ e^{j k s_0 \ve_3^{\tran}(\vq_k - \vq_l)}\nabla g(\vq_k - \vq_j, s_0) \Big),\label{eq:dZqp_dqq}
\end{align}
where the function $g(\vq_k - \vq_l, s_0)$ and its gradient $\nabla g(\vq_k - \vq_l, s_0)$ are available in \eqref{eq:g} and \eqref{eq:dg_ddelta} of Appendix~\ref{ap:A3}, respectively. Moreover, note that $\nabla Z_{kl} = -\nabla Z_{kl}$.
Finally, the position of the $k$-th \gls{ris} element is updated by solving the problem
\begin{problem}\label{eq:P5}
\begin{align}
         \displaystyle  \min_{\vq_k^{(i+1)}} &\|\vq_k^{(i+1)} -\vq_k^{(i)} - \alpha_i \nabla f(\Delta\vq_k)\|^2 \\
         & \mathrm{s.t.} \quad \vq_k^{(i+1)}\in\mathcal{Q}\\
         & \quad \quad \,\, \vq_k^{(i+1)}\, : \, \|\mG^{(i)}\Deltab^{(k)}_{\rmS\rmS}\|\ll 1,
\end{align}    
\end{problem}
\noindent where $\nabla f(\Delta\vq_k)$ is obtained by substituting \eqref{eq:dZqp_dqq} into \eqref{eq:partial_h} and \eqref{eq:partial_hconj}, which are in turn substituted into \eqref{eq:nabla_f}.

In Problem~\ref{eq:P5}, the second constraint is again needed to ensure convergence of the Neumann series. In particular, this latter constraint is imposed by operating a line search on the maximum value of $\alpha_i$, which ensures that the update of the $k$-th position of the \gls{ris} element is small enough to reach convergence of the Neumann series. Moreover, in order to guarantee convergence of the overall algorithm, we also impose a monotonic increase of the objective function in Problem~\ref{eq:P4}, which is achieved by ensuring that $f(\vq_k^{i+1})\geq f(\vq_k^{i})$.

In the following, we describe several possible strategies to define the feasible set $\mathcal{Q}$ for the positions of the \gls{ris} elements.

\subsection{3D sphere} \label{subsec:3Dsphere}

As a first example, we consider a \gls{3d} sphere of given radius $R$. Hence, $\mathcal{Q}$ is modeled as
\begin{align}
    \mathcal{Q} = \{\vq\in\Real^{3\times 1}\,:\,\|\vq\| \leq R\},
\end{align}
where the elements of $\vq_k= \begin{bmatrix}
    x_k \!\!&\!\! y_k \!\!&\!\! z_k
\end{bmatrix}^{\tran}$ represent the Cartesian coordinates of the $k$-th \gls{ris} element.

Hence, in this case, the projection operation is achieved as
\begin{align}
    P_{\mathcal{Q}}(\vq) = \begin{cases}
        \vq & \text{if } \|\vq\|\leq R\\
        \frac{\vq}{\|\vq\|} R & \text{otherwise}
    \end{cases}.
\end{align}

\subsection{Planar array}
In this case, the projection operation is achieved by fixing either the first or second component of each $\vq_k$ in Cartesian coordinates to a specified value. Indeed, the former case corresponds to the \gls{ris} laying on the $y-z$ plane, while the latter is associated with the \gls{ris} laying on the $x-z$ plane. Moreover, the remaining coordinates are given within specified limits (e.g., in the case of an \gls{ris} on the $y-z$ plane) as
\begin{align}
    & y_{\mathrm{min}} \leq y_k \leq y_{\mathrm{max}}, \quad z_{\mathrm{min}} \leq z_k \leq z_{\mathrm{max}} \quad \forall k.
\end{align}
Let $x$ denote the fixed coordinate of the planar array, therefore, we modify Problem \ref{eq:P5} as
\begin{problem}\label{eq:P3_planar}
\begin{align}
    \begin{array}{cl}
         \displaystyle \min_{\vq_k^{(i+1)}} & \|\vq_k^{(i+1)}-\vq_k^{(i)} + \alpha_i \nabla f(\Delta\vq_k)\|^2\\
         \mathrm{s.t.}
         & x_k^{(i+1)} = x\quad \forall k\\
         & y_{\mathrm{min}}\leq y_k^{(i+1)} \leq y_{\mathrm{max}}\quad \forall k\\
         & z_{\mathrm{min}}\leq z_k^{(i+1)} \leq z_{\mathrm{max}}\quad \forall k\\
         & \vq_k^{(i+1)}\, : \, \|\mG^{(i)}\Deltab^{(k)}_{\rmS\rmS}\|\ll 1.
    \end{array}
\end{align}    
\end{problem}

\subsection{Spherical array}

In this case, we express each vector $\vq_k$ in a spherical coordinate system as
\begin{align}\label{eq:q_k_spherical}
    \vq_k = \begin{bmatrix}
        \rho_k &
        \theta_k &
        \phi_k
    \end{bmatrix}^{\tran} \forall k,
\end{align}
where $\rho_k$ represents the distance from the \gls{ris} reference point, while $\theta_k$ and $\phi_k$ represent the azimuth and elevation angles, respectively. Moreover, we have that 
\change{
    $\rho_k = \sqrt{x_k^2 + y_k^2 + z_k^2}$, $\theta_k = \arctan\frac{y_k}{x_k}$, and $\phi_k = \arccos\frac{z_k}{\rho_k}$.}
Accordingly, the gradient of the objective function needs to be re-scaled as
\begin{align}
    \nabla f(\Delta\vq_k) = \begin{bmatrix}
    \frac{d\,f}{d\,\rho_k} & \frac{1}{\rho_k \sin \phi_k}\frac{d\,f}{d\,\theta_k} & \frac{1}{\rho_k}\frac{d\,f}{d\,\phi_k}    
    \end{bmatrix}^{\tran}.
\end{align}

Let $R$ denote the chosen radius, we thus modify Problem~\ref{eq:P5} as follows
\begin{problem}\label{eq:P3_spher}
\begin{align}
    \begin{array}{cl}
         \displaystyle \min_{\vq_k^{(i+1)}} & \|\vq_k^{(i+1)}-\vq_k^{(i)} + \alpha_i \nabla f(\Delta\vq_k)\|^2\\
         \mathrm{s.t.} 
         & \rho_k^{(i+1)} = R\quad \forall k\\
         & \theta_{\mathrm{min}}\leq \theta_k^{(i+1)} \leq \theta_{\mathrm{max}}\\
         & \phi_{\mathrm{min}}\leq \phi_k^{(i+1)} \leq \phi_{\mathrm{max}}\\
         & \vq_k^{(i+1)}\, : \, \|\mG^{(i)}\Deltab^{(k)}_{\rmS\rmS}\|\ll 1.
    \end{array}
\end{align}    
\end{problem}

\subsection{Cylindrical array}\label{subsec:cyl}

In this case, we modify the formulation in~\eqref{eq:q_k_spherical} as
\change{$\rho_k = \sqrt{x_k^2 + y_k^2}$, $\theta_k = \arctan\frac{y_k}{x_k}$, and $\phi_k = z_k$,}
where $z_k$ represents the height of the location of the $k$-th \gls{ris} element. Similarly, as in the spherical array case, the gradient of the objective function needs to be re-scaled as
\begin{align}
    \nabla f(\Delta\vq_k) = \begin{bmatrix}
    \frac{d\,f}{d\,\rho_k} & \frac{1}{\rho_k }\frac{d\,f}{d\,\theta_k} & \frac{d\,f}{d\,\phi_k}    
    \end{bmatrix}^{\tran}.
\end{align} 

\begin{figure}
    \centering
    \includegraphics[width=\linewidth]{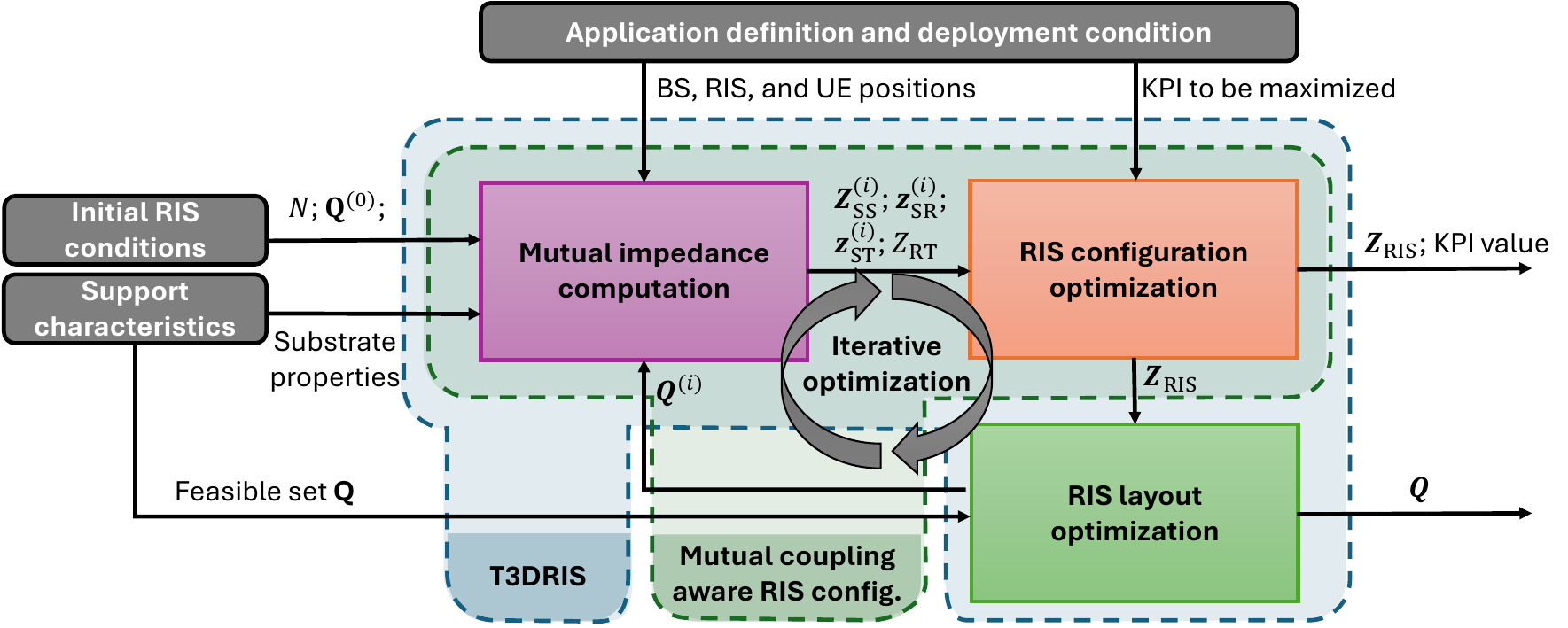}
    \caption{Graphical description of the proposed \name{} approach.}
    \label{fig:block_scheme}
\end{figure}

The complete formalization of the proposed optimization framework is provided in Algorithm~\ref{alg:A1}, while Fig.~\ref{fig:block_scheme} provides a graphical view of its corresponding block diagram. We remark here that appropriate modelling of the set of feasible \gls{ris} element positions $\mathcal{Q}$ opens the possibility to employ the proposed \name{} framework not only during the \gls{ris} design stage, but also \emph{a posteriori}, i.e., for the optimization of existing antenna systems and their inherently flexible geometry (e.g., smart skins~\cite{SmartSkins}, fluid antennas~\cite{FluidAntennas}, and movable antenna systems~\cite{MovableAntennas}).

\subsection{Convergence analysis}\label{subsec:CA}

Our proposed approach consists of an alternating optimization between the two optimization variables, namely the \gls{ris} impedance matrix $\mZ_{\rmRIS}$ and the \gls{3d} position of each of its elements $\mQ$, which are disjoint thanks to the structure of Problem~\ref{eq:P1}. Hence, we can decouple our optimization by alternating between the solution of Problem~\ref{eq:P2} and Problem~\ref{eq:P3}. The former is optimized via the approach in~\cite{Hassani23}, which is shown to converge to a stationary point of Problem~\ref{eq:P2} with a monotonically increasing behavior of its objective function. Whereas, for a given $\mZ_{\rmRIS}$ the optimization of $\mQ$ is given in Problem~\ref{eq:P3}. Firstly, our proposed approach proceeds by applying the Neumann series approximation for each $k$-th \gls{ris} element, thus obtaining Problem~\ref{eq:P4}. Secondly, we apply a projected gradient ascent~\cite{Bertsekas}, where the projection onto the feasible set $\mathcal{Q}$ is detailed in Problem~\ref{eq:P5}.
At each iteration of the algorithm, we carefully select the step size $\alpha_i$ such that the following conditions are met: \textit{i)} convergence of the Neumann series, which is guaranteed by the condition $\|\mG^{(i)}\Deltab^{(k)}_{\rmS\rmS}\|\ll 1$, \textit{ii)} monotonic increase of the objective function in Problem~\ref{eq:P4}, which is achieved by ensuring that $f(\vq_k^{i+1})\geq f(\vq_k^{i})$, with equality implying that $\vq_k^{i}$ is a stationary point of Problem~\ref{eq:P4}, and \textit{iii)} feasibility of the solution, which is ensured by considering only an updated position of each $k$-th \gls{ris} element belonging to the feasible set $\mathcal{Q}$. 
The aforementioned choice of the step size $\alpha_i$ together with the assumption on the convexity of $\mathcal{Q}$ guarantees convergence to a stationary point of Problem~\ref{eq:P4}~\cite{Bertsekas}. 

Therefore, starting from the solution of Problem~\ref{eq:P2} with its optimized $\mZ_{\rmRIS}$, the projected gradient ascent can never decrease the objective function thanks to the aforementioned choice of $\alpha_i$. Lastly, we note that the objective function of Problem~\ref{eq:P1} is bounded from above thanks to the physical nature of the problem, i.e., the passive nature of the \gls{ris}, and since the location of the \gls{ris} elements is bounded. Finally, we thus conclude that every limit point of the sequence of solutions in $\mZ_{\rmRIS}$ and $\mathcal{Q}$ obtained with the proposed approach is a stationary point of the original Problem~\ref{eq:P1}.

\begin{algorithm}[t!]
  \caption{\name{}}\label{alg:A1}
  \begin{algorithmic}[1]
     \State Initialize $\epsilon$, $\mathrm{I}_{\mathrm{max}}$, $\mathrm{SNR}^{(0)}$, $i=0$, and $\mQ^{(0)}$ 
     \State Given $\mQ^{(0)}$, calculate all $Z_{qp}$ by using \eqref{eq:zqp_compact} 
     \While {$\big(\mathrm{SNR}^{(i)}-\mathrm{SNR}^{(i+1)}\big)/\mathrm{SNR}^{(i)}>\epsilon$ $\wedge$ $i<\mathrm{I}_{\mathrm{max}}$} 
     \State For a fixed $\mQ^{(i)}$, optimize $\mZ_{\rmRIS}$ using~\cite{Hassani23} 
     \For {$k=1,\ldots,N$}
     \State Calculate $\nabla f(\Delta\vq_k)$ by substituting \eqref{eq:partial_delta_Z} into \eqref{eq:partial_h} and \eqref{eq:partial_hconj}, which are in turn substituted in \eqref{eq:nabla_f} 
     \State Find the maximum value of $\alpha_i$ that ensures $\|\mG^{(i)}\Deltab^{(k)}_{\rmS\rmS}\|\ll 1$ \change{and $f(\vq_k^{i+1})\geq f(\vq_k^{i})$} via bisection
     \State Set $\vq_k^{i+1}$ as the solution of Problem~\ref{eq:P5} 
     \State Recalculate all $Z_{qp}$ by using \eqref{eq:zqp_compact} 
     \EndFor
     \State $i=i+1$
     \EndWhile
     \State Output: $\mQ$ and $\mZ_{\rmRIS}$
  \end{algorithmic}
\end{algorithm}

\subsection{Computational complexity}\label{subsec:CC}

The overall computational complexity of the proposed \name{} approach is given by the number of iterations necessary for convergence, i.e., $\mathrm{I}_{\mathrm{conv}}$, times the sum of the computational complexity of the algorithm in~\cite{Hassani23}, which is used for the optimization of the \gls{ris} configuration, and the computational complexity of the \gls{3d} position update of each \gls{ris} unit cell via iterative Neumann series and projected gradient ascent. In our single-antenna \gls{ue} setup, the former is given by $\mathcal{O}(N^4+N^3)$, where $N$ is the number of \gls{ris} elements~\cite{Hassani23}. Whereas, the complexity of the \gls{3d} \gls{ris} element positions is dictated essentially by the calculation of the gradient $\nabla f (\Delta \vq_k)$, $k=1,\ldots,N$, whose complexity is $\mathcal{O}(N)$, the bisection search for the maximum value of $\alpha_i$, which requires $\gamma^{\scriptscriptstyle \mathrm{bis}}$ iterations, with $\gamma^{\scriptscriptstyle \mathrm{bis}}\leq \ceil*{\log_2(\frac{1}{\epsilon^{\scriptscriptstyle \mathrm{bis}}})}$ and $\epsilon^{\scriptscriptstyle \mathrm{bis}}$ the tolerance used to reach convergence, and the solution of Problem~\ref{eq:P5}, which essentially boils down to the projection onto the feasible set $\mathcal{Q}$ and is again $\mathcal{O}(N)$ in the considered scenarios. Lastly, all mutual impedances $Z_{qp}$ are re-calculated for every update of the \gls{ris} unit cells. In particular, for every update of each $\vq_k$, the mutual impedance with the \gls{bs}, \gls{ue} and the remaining $N-1$ \gls{ris} unit cells need to be re-calculated. The overall complexity of \name{} is thus given by $\mathcal{O}(\mathrm{I}_{\mathrm{conv}}(N^4+N^3+ N(\gamma^{\scriptscriptstyle \mathrm{bis}}+N+3)))$.

We remark here that, while the computational complexity of \name{} may increase significantly with the number of \gls{ris} elements, the proposed approach is intended to be applied offline during the network planning phase. Indeed, the optimization of the \gls{ris} shape needs to be carried out before its fabrication and is not intended to be changed dynamically.

\begin{figure}
     \centering
     \hspace{-0.45cm}
     \subfloat[\scriptsize ULA, $\lambda/16$ inter-el. spacing.]{
         \includegraphics[width=0.24\columnwidth,height = 0.17\columnwidth]
         {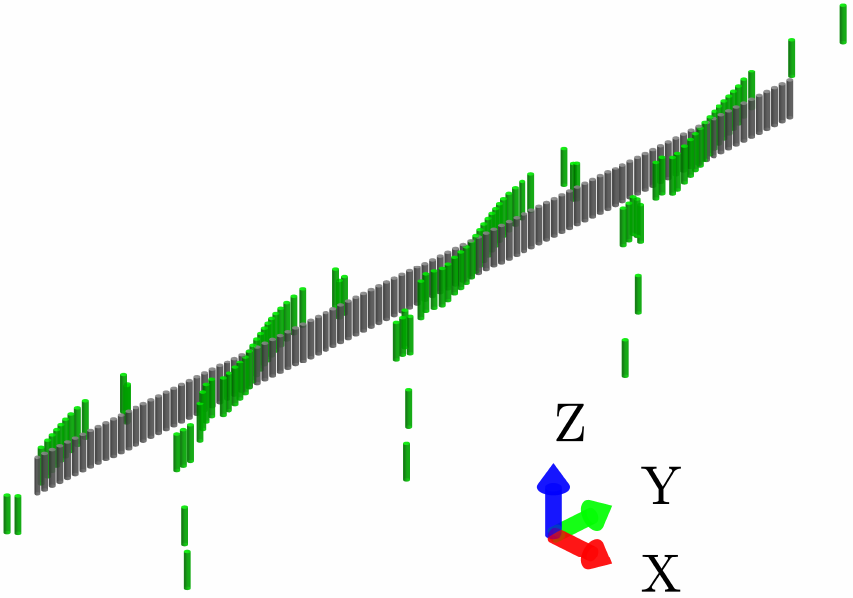}
     \hspace{-0.1cm}
     \includegraphics[height = 0.2\columnwidth]
     {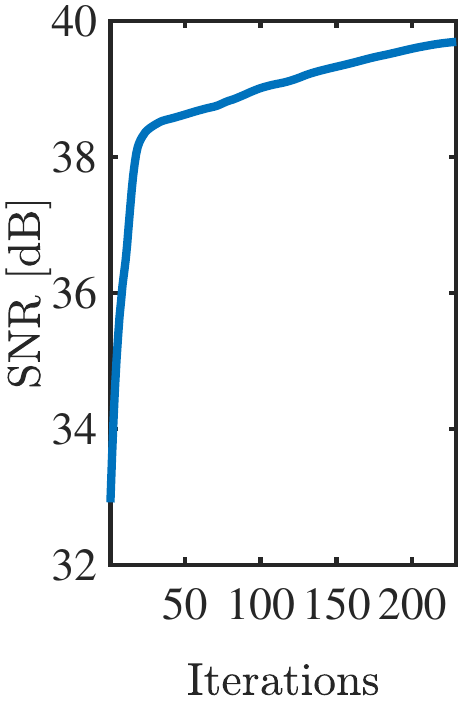}}
    \hspace{1cm}
     \subfloat[\scriptsize UPA, $\lambda/2$ inter-el. spacing.]{
         \includegraphics[width=0.22\columnwidth,height = 0.2\columnwidth]
         {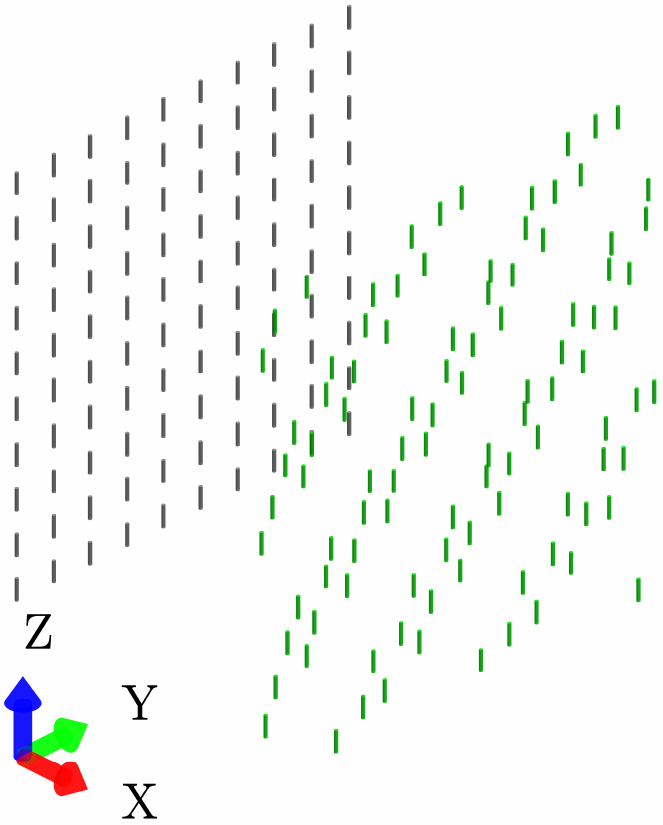}
     \hspace{0.25cm}
     \includegraphics[height = 0.2\columnwidth]
     {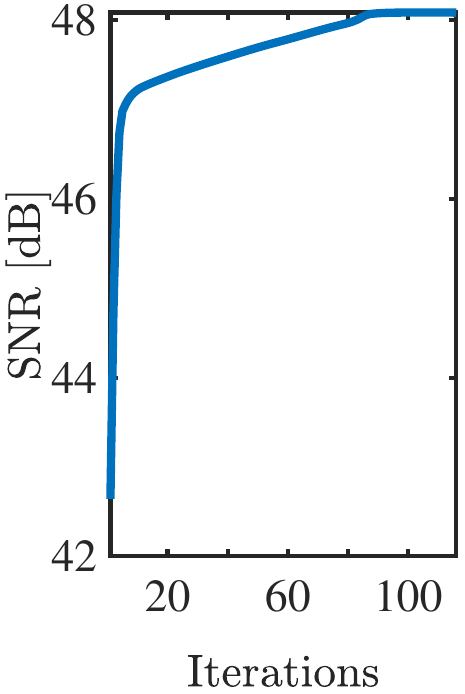}}
     \\
\subfloat[\scriptsize Cylindrical, $\lambda/2$ inter-el. spacing and radius $0.1$~m.]{\includegraphics[width=0.2\columnwidth,height = 0.24\columnwidth]
{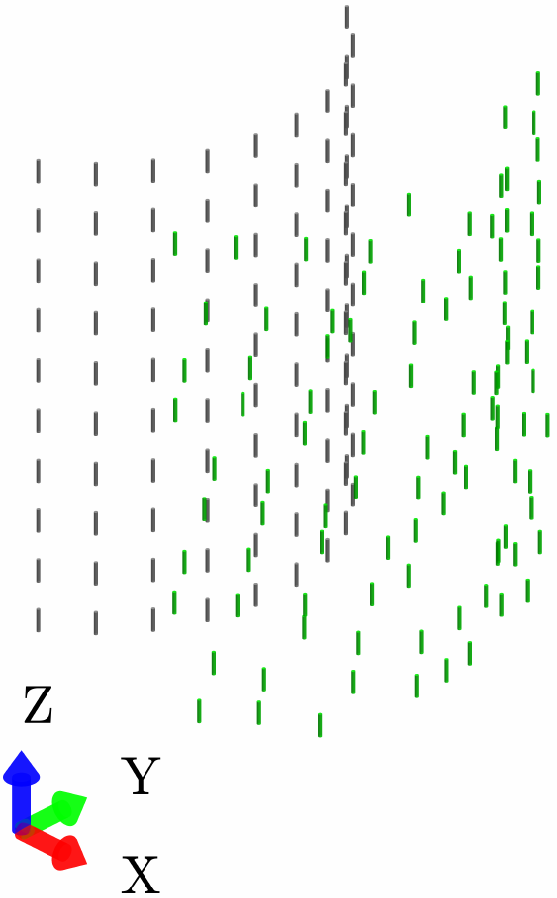}
\hspace{0.3cm}
\includegraphics[height = 0.23\columnwidth]
     {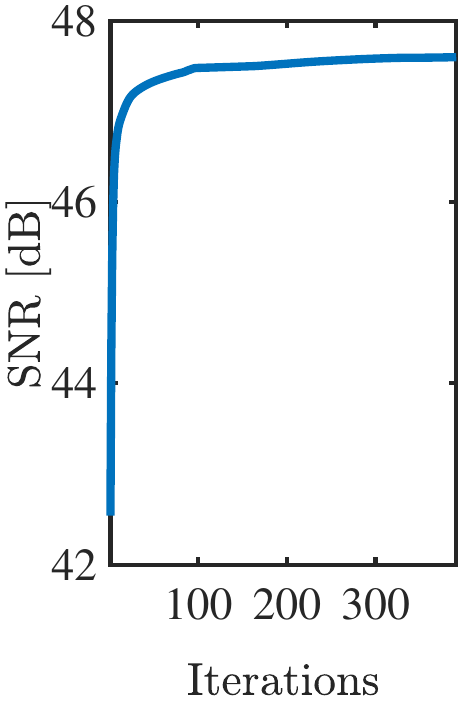}}
     \hspace{1cm}
     \subfloat[\scriptsize Spherical, $\lambda/2$ inter-el. spacing and radius $0.1$~m.]{\includegraphics[width=0.22\columnwidth,height = 0.2\columnwidth]
     {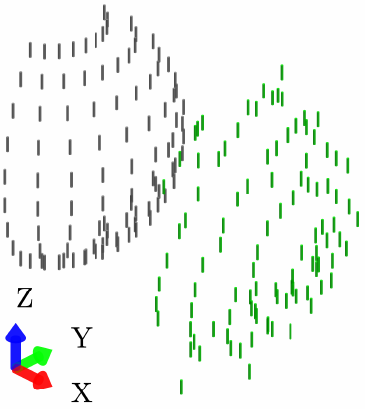}
     \hspace{0.3cm}
    \includegraphics[height = 0.23\columnwidth]
     {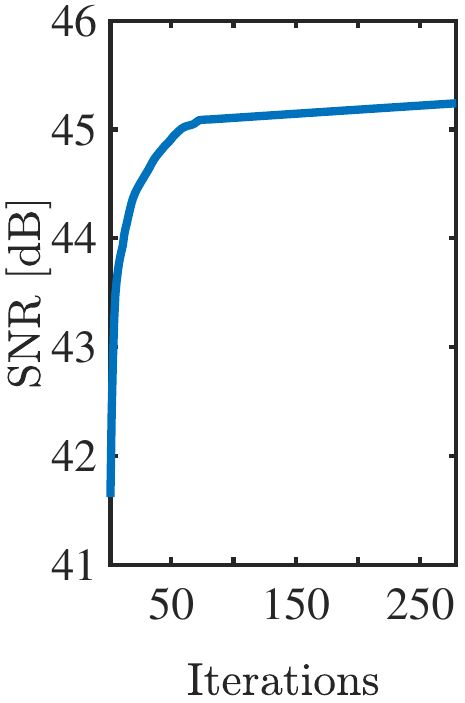}}
        \caption{Four different \gls{ris} structures (black color) with their corresponding non-planar element arrangements, optimally designed with \name{} in the unconstrained case (green color), whose performance in terms of \gls{snr} versus algorithm iterations is shown on the right-hand side.}
        \label{fig:structures}
\end{figure}

\section{Performance evaluation}\label{sec:results}

We consider a scenario wherein the center-point of the \gls{ris} is located at the origin, while the single-antenna \gls{bs} and \gls{ue} are located at $[1.3,0,0]~$m and $[0.98,0.56,-0.65]~$m, respectively, which correspond to a distance of $1.3$~m and azimuth of elevation angles of $30^\circ$, $120^\circ$ from the perspective of the \gls{ris}, respectively, while the transmit signal from the \gls{bs} is incident from the broadside of the \gls{ris}. We consider a total of $N = 100$ $\lambda/2$-length dipoles forming the \gls{ris}, unless otherwise stated.\footnote{In conventional \gls{ris} designs, the distance between the element centers is kept at $\lambda/2$ while the size of the elements themselves, such as patch antennas, must be slightly smaller to avoid superposition or contact.} The initial positions $\mQ^{(0)}$ of the dipoles follow a regular conformal structure. In particular, we consider four specific structures, $i$) \gls{ula} with $\lambda/16$ inter-element spacing\footnote{The setup at $\lambda/16$ aims at keeping the overall dimension of the \gls{ula} proportional with the other initial structures considered, which results in a comparable average pathloss among \gls{ris} elements and the \gls{bs}/\gls{ue} in all the experiments.}, $ii$) \gls{upa} with $\lambda/2$ inter-element spacing, $iii$) cylindrical \gls{ris} with $\lambda/2$ inter-element spacing and radius $0.1$~m, and $iv$) spherical \gls{ris} with $\lambda/2$ inter-element spacing and radius $0.1$~m, as shown in Fig.~\ref{fig:structures}. Moreover, for each of the specific initial \gls{ris} structures, we distinguish two cases: \emph{i}) unconstrained and \emph{ii)} constrained optimization. In the former case, the feasible set for the positions of the \gls{ris} elements is a \gls{3d} sphere of radius $R=0.05$~m centered at the origin, whereas in the latter case, the optimized \gls{ris} shape must preserve the general geometrical shape of the initial position, which are chosen having in mind practical constraints on \gls{ris} fabrication (e.g., cylinder of a given radius, a plane with given maximum height and width, etc.).
\noindent In this case, the limits on the Cartesian/polar coordinates are given as $1.5$ times the ones corresponding to the initial \gls{ris} shape. The operating frequency is set to $30$ GHz, i.e., $\lambda=1$~cm, while we set $R_0 = 0.2$ and $Y_0=1$ without loss of generality. The transmit power at the \gls{bs} and noise power are set as $P = 10$~dBm and $\sigma_n^2 = -80$~dBm, respectively. In all the experiments, the maximum number of iterations and the convergence threshold are set to $I_{\scriptscriptstyle \mathrm{max}} = 2000$ and $\epsilon = 10^{-6}$, respectively. Lastly, the set of feasible \gls{ris} impedances is given by $\mathcal{B} = [-5000,188]~\Omega$. The relevant simulation parameters are summarized in Table~\ref{tab:params}.
\begin{table}[h!]
\caption{Simulation parameters.}
\label{tab:params}
\centering
\resizebox{\linewidth}{!}{%
\renewcommand{\arraystretch}{1}
\begin{tabular}{|c|c|c|c|}
\hline
\cellcolor[HTML]{EFEFEF} \textbf{Parameter} & \textbf{Value} &\cellcolor[HTML]{EFEFEF} \textbf{Parameter} & \textbf{Value}\\
\hline
 \cellcolor[HTML]{EFEFEF}  $\vp_{\scriptscriptstyle \mathrm{BS}}$  & $[1.3,~0,~0]$~m & \cellcolor[HTML]{EFEFEF} $\vp_{\scriptscriptstyle \mathrm{UE}}$ &  $[0.98,~0.56,~-0.65]$~m \\
\hline 
 \cellcolor[HTML]{EFEFEF} $\mathrm{R}$ & $0.05$~m &\cellcolor[HTML]{EFEFEF} $N$ & $100$  \\
\hline 
\cellcolor[HTML]{EFEFEF} $\rm{R}_0$ & $0.2~\Omega$  &
\cellcolor[HTML]{EFEFEF} $P$ & $10$~dBm  \\
\hline
\cellcolor[HTML]{EFEFEF} $I_{\scriptscriptstyle \mathrm{max}}$ & $2000$  &
\cellcolor[HTML]{EFEFEF} $\epsilon$ & $10^{-6}$ \\
\hline
\cellcolor[HTML]{EFEFEF} $\lambda$  & $1$~cm & \cellcolor[HTML]{EFEFEF} $Y_{0}$ & $1$ \\
\hline
\cellcolor[HTML]{EFEFEF} $\sigma_n^2$ & $-80$~dBm & \cellcolor[HTML]{EFEFEF} $\mathcal{B}$ & $[-5000,188]~\Omega$\\
\hline
\end{tabular}
}
\renewcommand{\arraystretch}{1}
\vspace{-3mm}
\end{table}

\begin{figure}
     \centering
     \hspace{-0.5cm}
     \subfloat[\scriptsize ULA, $\lambda/16$ inter-el. spacing.]{
         \includegraphics[width=0.25\columnwidth,height = 0.15\columnwidth]
         {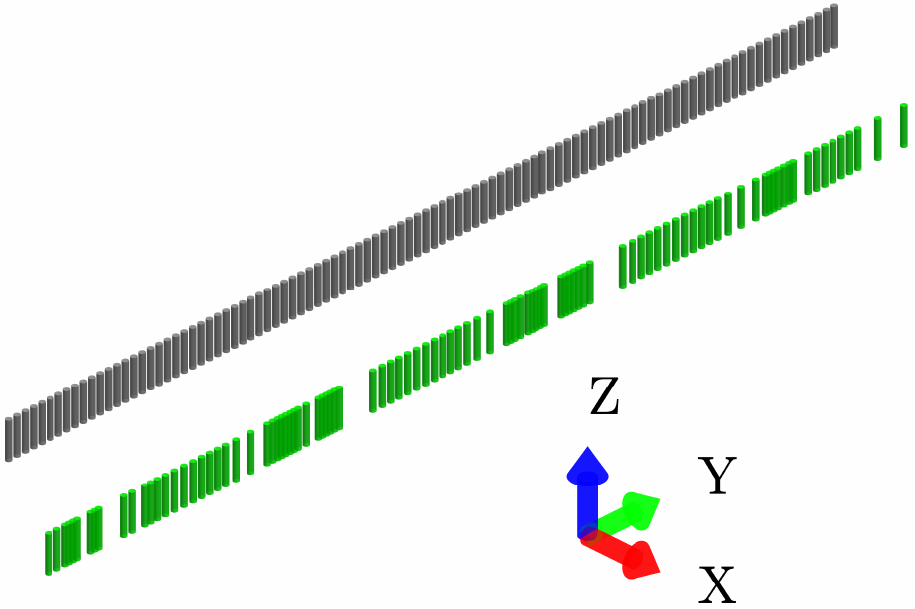}
     \includegraphics[height = 0.23\columnwidth]
     {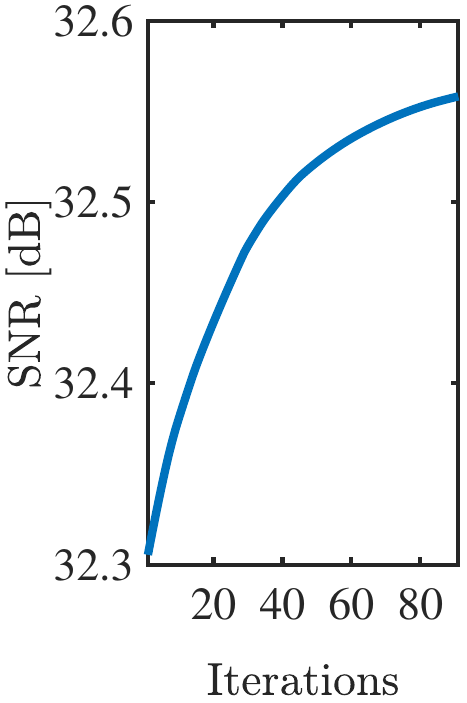}}
    \hspace{0.5cm}
     \subfloat[\scriptsize \gls{upa}, $\lambda/2$ inter-el. spacing.]{
         \includegraphics[width=0.21\columnwidth,height = 0.21\columnwidth]
         {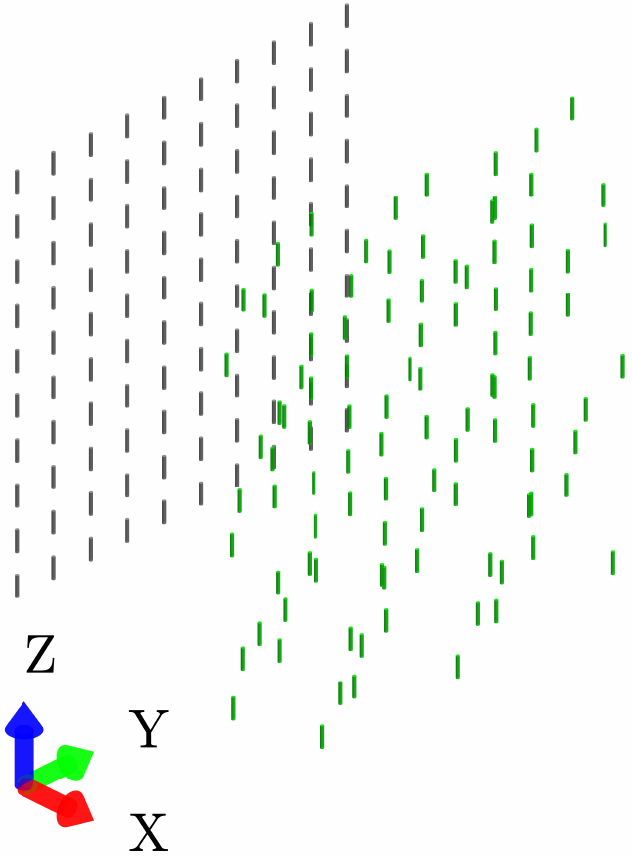}
     \hspace{0.3cm}
     \includegraphics[height = 0.23\columnwidth]
     {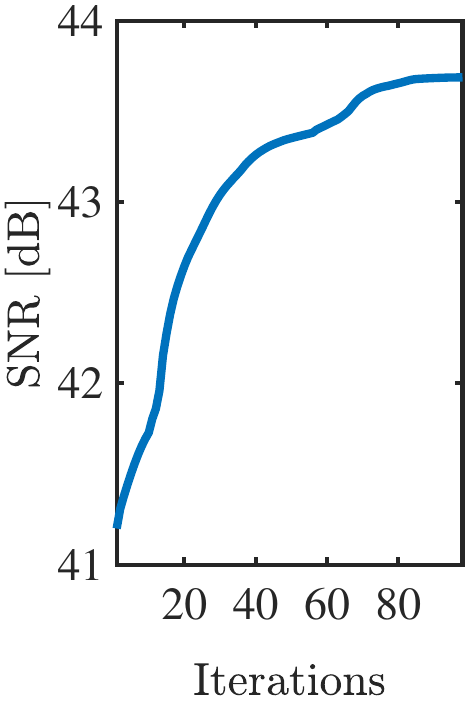}}
     \\
\subfloat[\scriptsize cylindrical, $\lambda/2$ inter-el. spacing and radius $0.1$~m.]{\includegraphics[width=0.19\columnwidth,height = 0.2\columnwidth]
{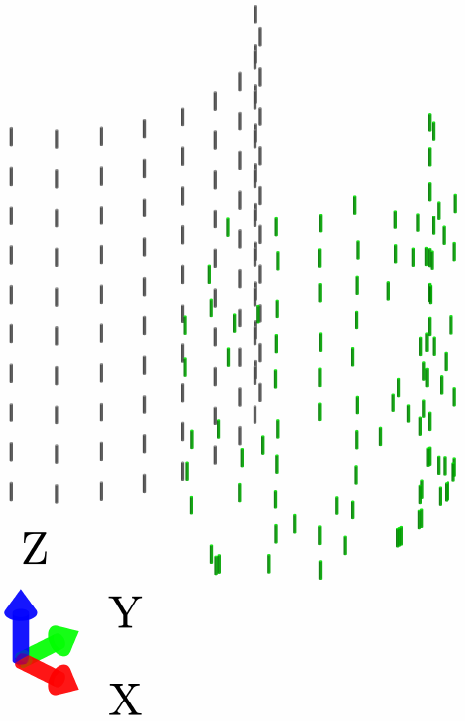}
\hspace{0.3cm}
\includegraphics[height = 0.2\columnwidth]
     {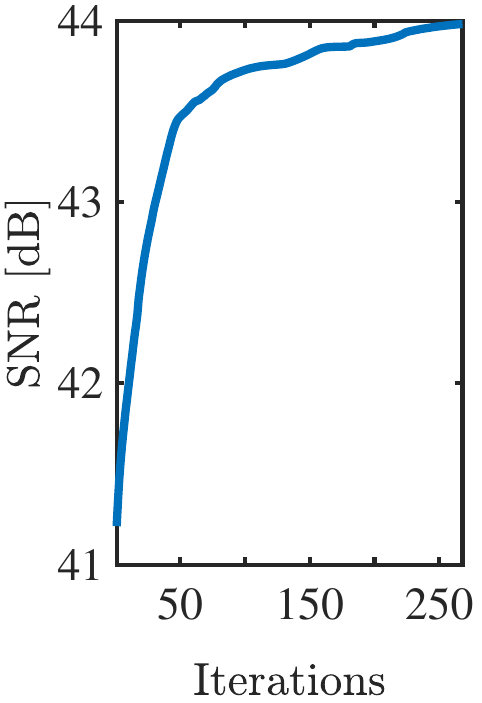}}
     \hspace{1cm}
     \subfloat[\scriptsize Spherical, $\lambda/2$ inter-el. spacing and radius $0.1$~m.]{\includegraphics[width=0.22\columnwidth,height = 0.2\columnwidth]
     {Figures/3D_surf_T3DRIS_Spher_v5.pdf}
     \hspace{0.2cm}
    \includegraphics[height = 0.2\columnwidth]
     {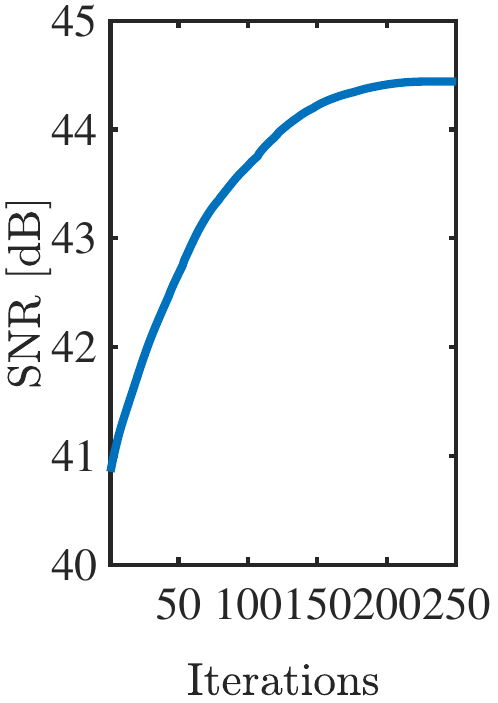}}
        \caption{Four different \gls{ris} structures (black color) with their corresponding non-planar element arrangements, optimally designed with \name{} in the constrained case (green color), whose performance in terms of \gls{snr} versus algorithm iterations is shown on the right-hand side.}
        \label{fig:structures_projected}
\end{figure}

\subsection{Numerical results}\label{subsec:NR}

\subsubsection{Unconstrained case}

Fig.~\ref{fig:structures} shows how the initial \gls{ris} structures can be optimally transformed within a 3D space thanks to the application of \name{}. For the four considered case studies, we can draw the following conclusions: $i$) the optimized \gls{ris} elements spread out at different heights along the $y$- and $z$-axis following regular \emph{striped} patterns, and $ii$) the \gls{ris} elements tend to move closer to the \gls{bs} and the \gls{ue}. Indeed, moving the positions of the \gls{ris} elements towards the \gls{bs} and \gls{ue} causes a reduction in the path loss between them, which ultimately increases \gls{snr}. Since our focus is mostly on the effects of the \gls{ris} shape, we chose the radius of the \gls{3d} sphere wherein the \gls{ris} elements can be freely moved small enough to limit such effects.

Interestingly, in the case of the \gls{ula}, the initial \gls{ris} structure is not able to steer energy at different elevation angles, hence the spread over the $z$-axis is more remarkable. It is noticeable that the optimized positions of the dipoles significantly increase the resulting \gls{snr}, hence empirically demonstrating that the optimization of the \gls{ris} structure is a viable approach to enhance communication performance.

As for the \gls{upa} geometry, we see that optimizing the position and spacing of the dipoles leads to a noticeable \gls{snr} improvement. It is also noticeable that, similarly to the previous example, the spread of the elements in space exhibits a regular and periodic pattern, with stripes that are spaced at regular intervals in the $x$-$y$ plane. 
Whereas, the cylindrical and spherical initial geometries tend to result in an optimized surface wherein the elements are concentrated along equally-spaced curved arches on the $x$-$y$ plane, which are more pronounced in the spherical case.

Overall, in all the considered scenarios we notice that the optimized \gls{ris} structure follows a certain periodic structure, which mimics the periodicity of the incoming and reflected plane waves.

\begin{figure}
    \centering
    \includegraphics[width=0.6\columnwidth]{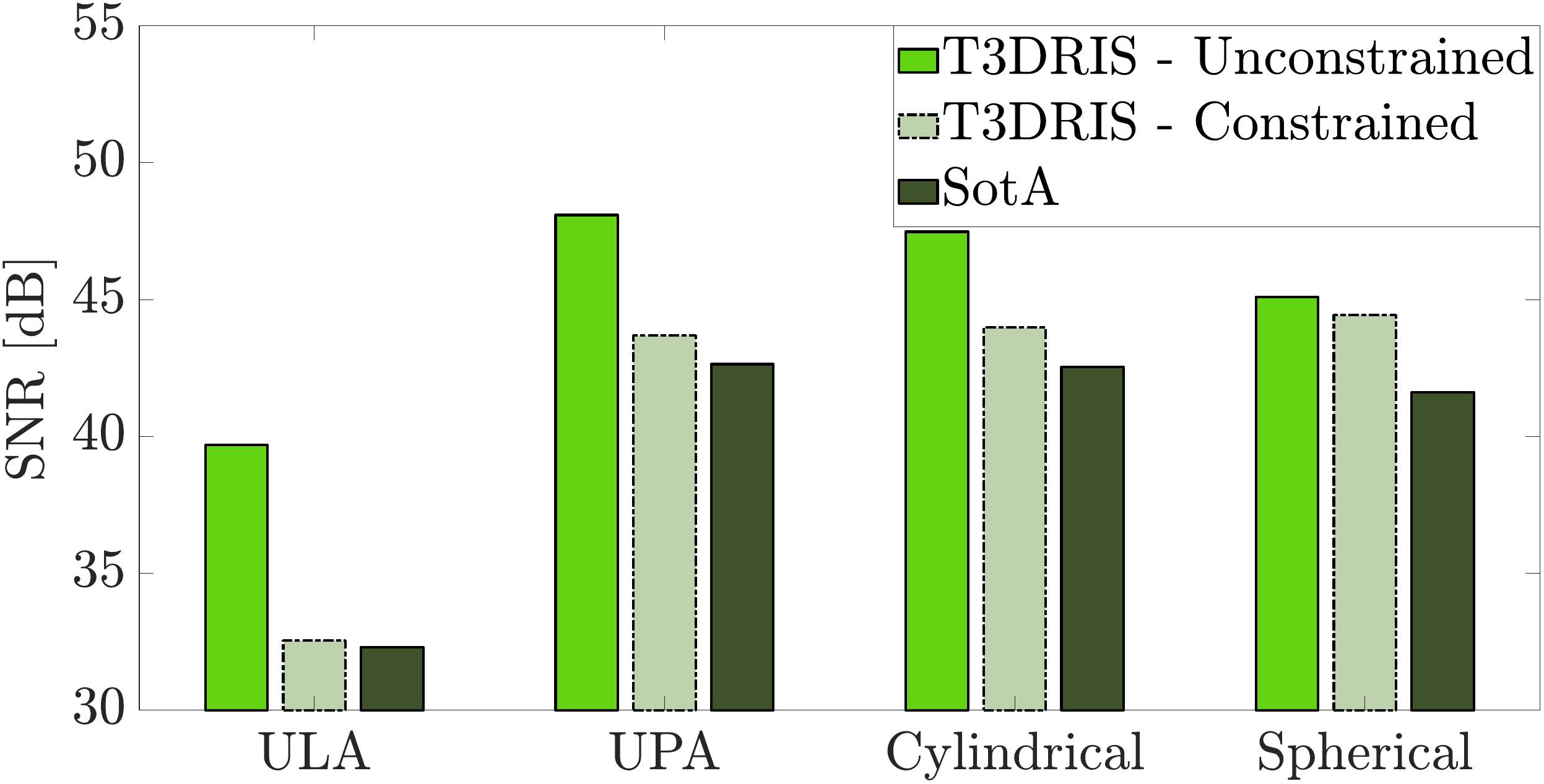}
    \caption{Bar plot of the maximum SNR obtained with the conventional and optimized \gls{ris} shapes.}
    \label{fig:SNR}
\end{figure}

\subsubsection{Constrained case} In Fig.~\ref{fig:structures_projected}, we demonstrate the effectiveness of our proposed \name{} approach in optimizing the element arrangements over a specific constrained shape. As for the unconstrained case, we notice that in all four considered scenarios, the optimized \gls{ris} element arrangements follow period patterns with non-uniform inter-element spacings. Moreover, we notice that compared with the unconstrained case, the \gls{snr} improvement is less pronounced, which is due to a reduction in the degrees of freedom in the set of feasible \gls{ris} element positions.

\begin{figure}
    \centering
     \subfloat[\scriptsize ULA.]{
         \includegraphics[width=0.46\columnwidth]
         {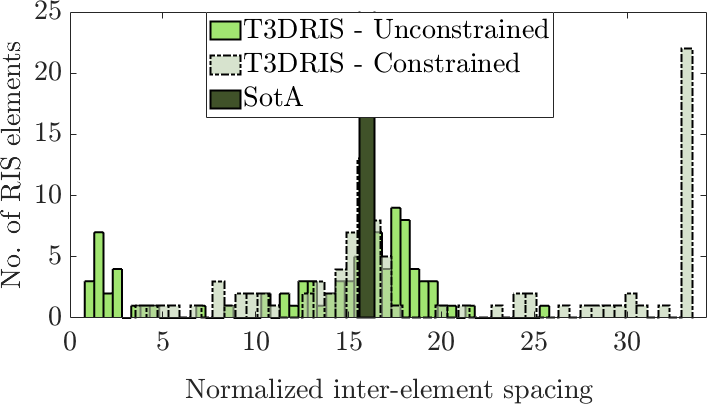}
     }
    \hspace{0.1cm}
     \subfloat[\scriptsize UPA.]{
         \includegraphics[width=0.46\columnwidth]
         {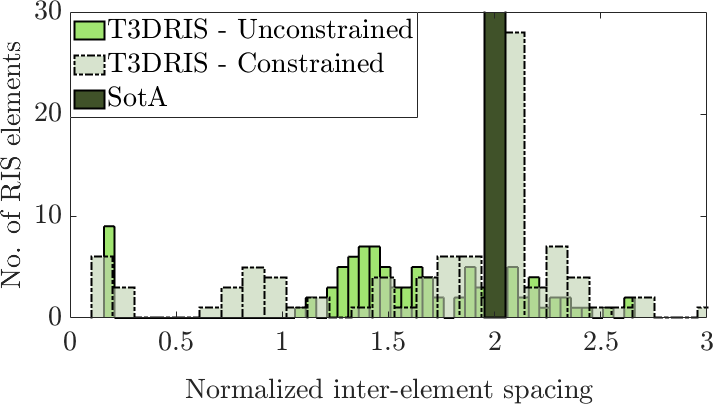}
     }
     \\
\subfloat[\scriptsize Cylindrical.]{\includegraphics[width=0.46\columnwidth]
{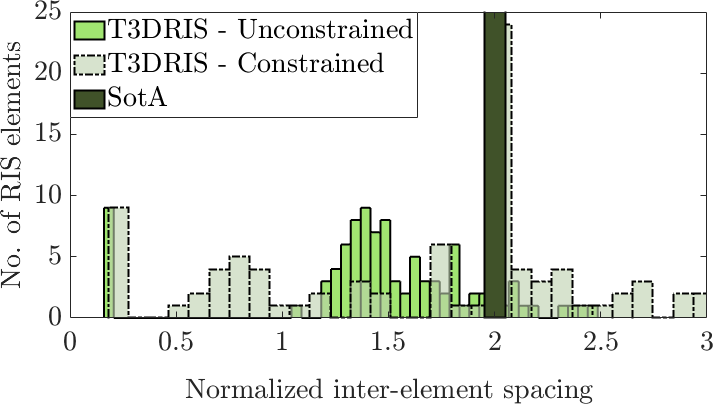}
}
     \hspace{0.2cm}
     \subfloat[\scriptsize Spherical.]{\includegraphics[width=0.46\columnwidth]
     {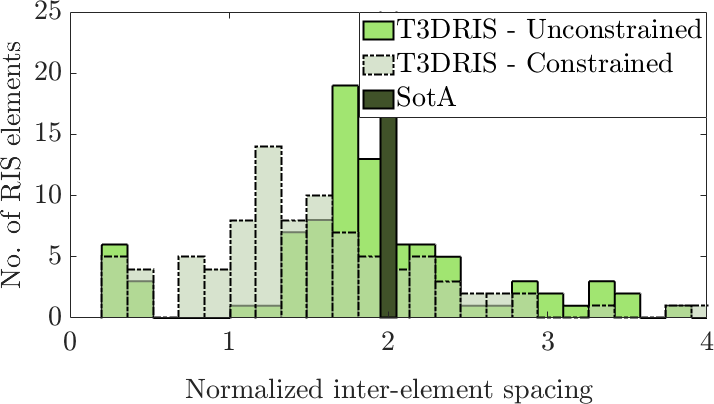}
     }
    \caption{Distribution of the normalized inter-element spacings ($90$th percentile), i.e., $\lambda/d_{ij}$ with $d_{qp}$ the distance between the \gls{ris} elements $q$ and $p$.}
    \label{fig:distribution}
\end{figure}

In Fig.~\ref{fig:SNR}, we show the initial \gls{snr} for each \gls{ris} structure (labeled as \emph{SotA}) and compare it against the \gls{snr} obtained after executing \name{} for both the unconstrained and constrained cases. Note that the initial \gls{snr} value corresponds to the performance of conventional  \gls{ris} devices, i.e., whose shape is not optimized for the given application scenario. Notably, the highest gain compared to the initial shape is obtained for the \gls{upa} geometry: this exacerbates the need to design metamorphic \glspl{ris} as compared to planar surfaces. In Fig.~\ref{fig:distribution}, we show the distribution of the normalized inter-element spacings at the $90$th percentile, i.e., $\lambda/d_{ij}$ with $d_{qp}$ the distance between elements $q$ and $p$, obtained with \name{} as compared to the initial shape.

\subsection{Full-wave electromagnetic simulations}

In this section, we provide full-wave \gls{em} simulations to validate our proposed approach against existing methods, using CST Studio Suite~\cite{cststudio}. 
In agreement with the proposed analytical model for conformal \glspl{ris}, the full-wave simulations are implemented by adopting a thin wire dipole as a unit cell. As for the numerical results in Section~\ref{subsec:NR}, the choice of freestanding dipoles is motivated by the: $i)$ availability of closed-form mathematical expressions for the mutual impedance (see~\eqref{eq:zqp_compact}), which facilitates the development of optimization algorithms, and allows for a deep understanding of the role of mutual coupling in the re-radiation pattern \cite{DiRenzo23,Russer2023}; $ii)$ freedom of arrangement through curved shells, i.e., as detailed in Section~\ref{sec:physics}, the resonant frequency of the unit cell does not depend on the radius of curvature of the \gls{ris}, similar to the case of planar structures~\cite{Vukovic2022}; 
In this regard, we compare our proposed approach against the method in~\cite{Mizmizi23}. Here, the authors design the optimized \gls{ris} configuration for a given cylindrical shape, by assuming a conventional geometric channel model based on steering vectors and ideal unitary reflection coefficients at the unit cells. However, such kind of models inherently assume that the impinging signal at the unit cell can be reflected with arbitrary delay and maximum efficiency, which in general is feasible only if the unit cell surface area is much larger than the signal wavelength. Moreover, the mutual coupling across the \gls{ris} board is not accounted for. Note that the relationship between reflection coefficients and load impedances at the \gls{ris} is not trivial and is still an open topic~\cite{nerini2023universal, abrardo2023design}. Therefore, in order to operate a fair comparison with our impedance-based model, we transform the \gls{ris} phase shifts obtained with the method in~\cite{Mizmizi23} $\{\theta_k\}_{k=1}^N$ into load impedances $\{Z_k\}_{k=1}^N$ as~\cite{Pozar}
\begin{align}
    e^{j\theta_k} = \frac{Z_{k}-Z_0}{Z_{k}+Z_0} \quad \forall k,\label{eq:rho_to_z}
\end{align}
where $Z_0$ is the characteristic (self) impedance of the \gls{ris} unit cell, which can be obtained with~\eqref{eq:zqp_compact}. In~\eqref{eq:rho_to_z}, the solution for each $Z_k$ leads to a complex number with both real and imaginary part. In order to have coherence with the \gls{ris} load impedance model in~\eqref{eq:Zris}, we fix the real part to the parameter $R_0$ and retain the imaginary part obtained by inverting~\eqref{eq:rho_to_z}.

\begin{figure}
     \centering
     \subfloat{
         \includegraphics[width=0.28\columnwidth,height = 0.2\columnwidth]{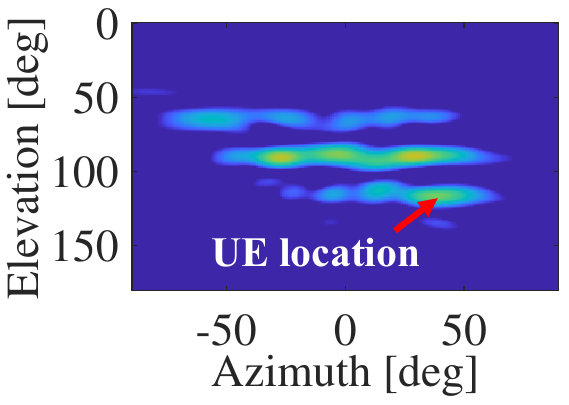}}
     \hspace{0.2cm}
     \subfloat{\includegraphics[width=0.28\columnwidth,height = 0.2\columnwidth]{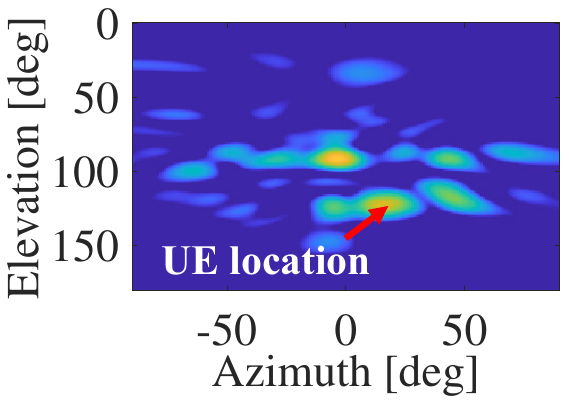}}
     \hspace{0.2cm}
     \subfloat{\includegraphics[width=0.3\columnwidth,height = 0.2\columnwidth]{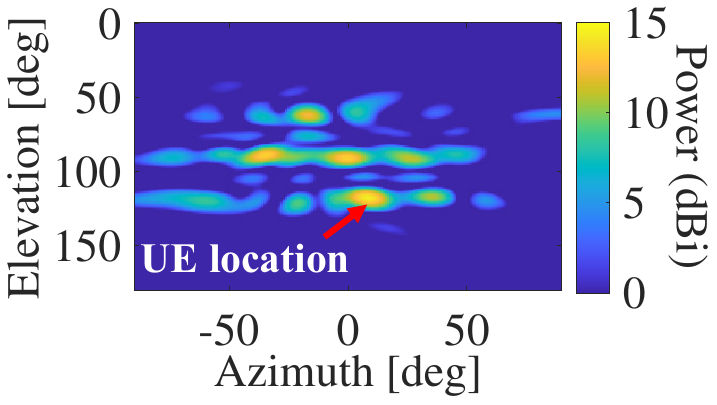}}
     \hspace{0.05cm}
     \subfloat[\scriptsize SotA algorithm in~\cite{Mizmizi23}.]{\setcounter{subfigure}{1}
         \includegraphics[width=0.29\columnwidth]{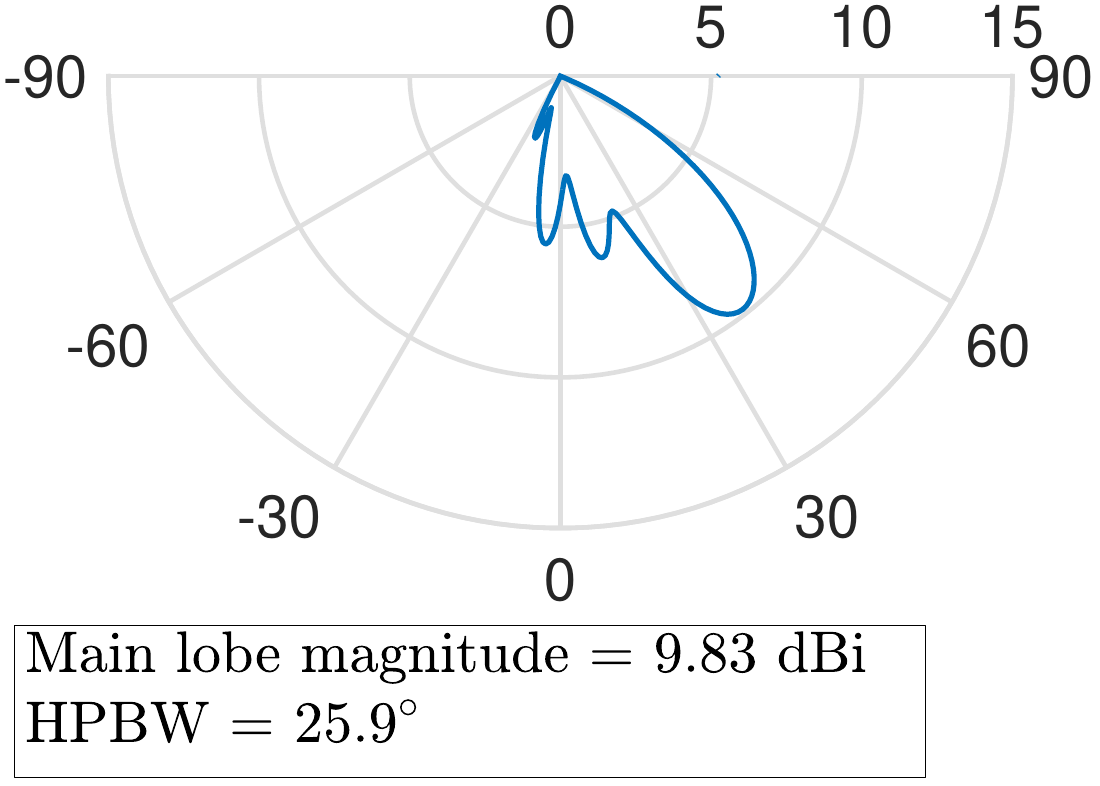}}
     \hspace{0.2cm}
     \subfloat[\scriptsize \name{} -\\ constrained case.]{\includegraphics[width=0.29\columnwidth]{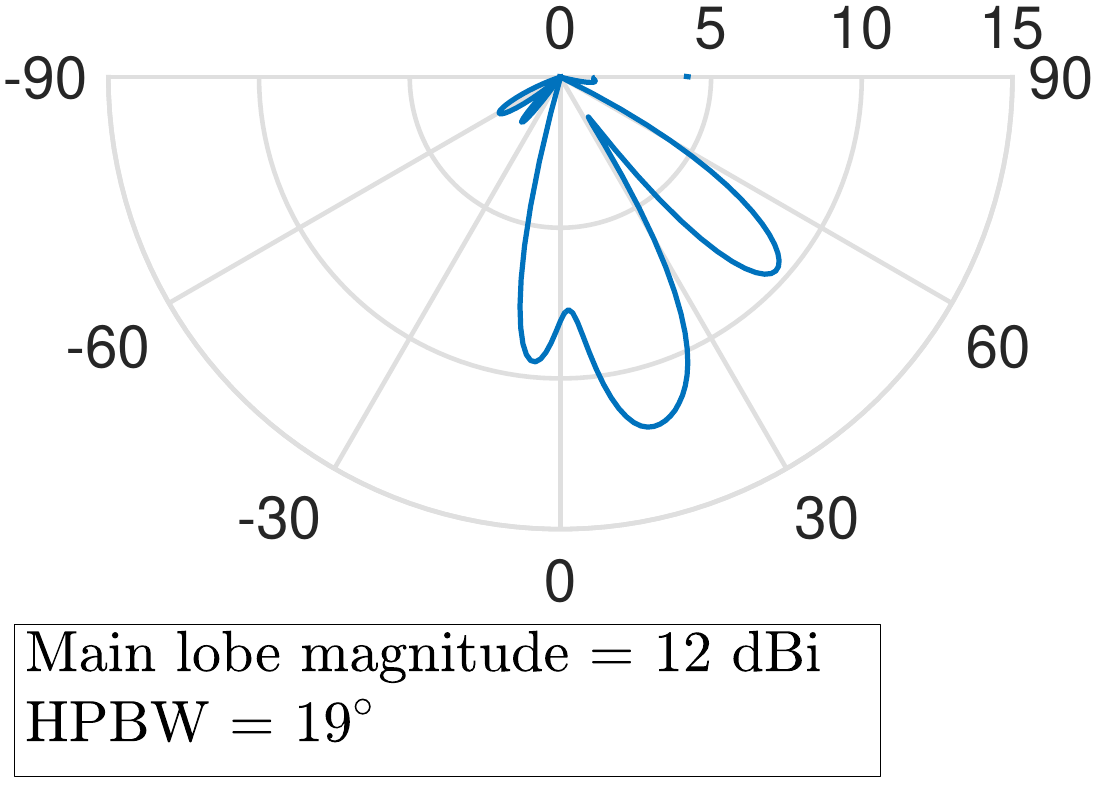}}
     \hspace{0.2cm}
     \subfloat[\scriptsize \name{} -\\ unconstrained case.]{\includegraphics[width=0.29\columnwidth]{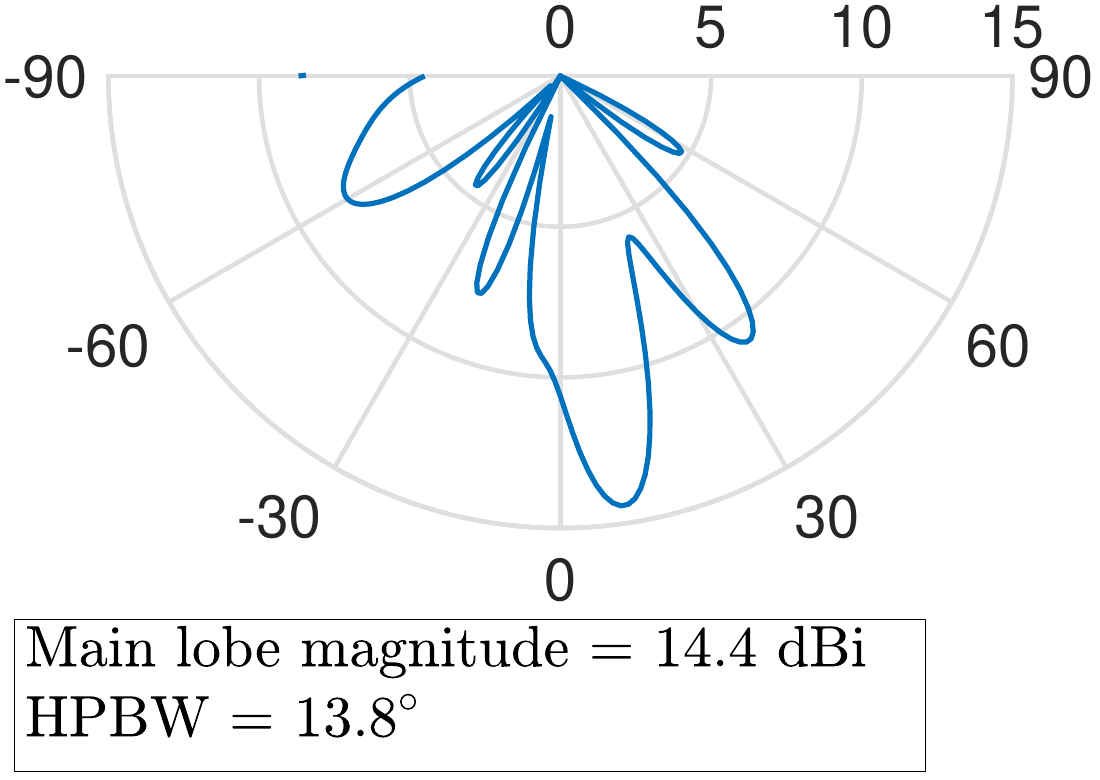}}     \caption{Heatmap of the full-wave simulation of the radiated power in azimuth and elevation at a cylindrical \gls{ris} of radius $0.1$~m originated by the conventional method~\cite{Mizmizi23}, and the optimized \gls{ris} shape and configuration via the proposed \name{} approach, and corresponding horizontal cuts at elevation angle of $120^\circ$.}
    \label{fig:CST}
\end{figure}

Fig.~\ref{fig:CST} shows the $2$D heatmap of the beampattern expressed as the power irradiated along the azimuth and elevation directions for the cylindrical \gls{ris} structure in Section~\ref{subsec:NR}, i.e., with radius $0.1$~m, which is optimized according to~\cite{Mizmizi23} and by using the proposed \name{} approach in both the unconstrained and constrained cases. We note that \name{} achieves a gain of $5$~dBi and $3.5$~dBi for the two cases, respectively, in terms of radiated power towards the location of the \gls{ue}, which is highlighted with a red arrow, as compared to the scheme proposed in~\cite{Mizmizi23}. Also, we show the polar plots of horizontal cuts at elevation angle of $120^\circ$, thus demonstrating that \name{} provides a narrower beamwidth, expressed in terms of the \gls{hpbw}, and thus a higher directivity, thanks to the employed \gls{em}-compliant and mutual coupling aware \gls{ris} model, and the joint optimization of the \gls{ris} shape and configuration. It is worth noting that the pointing direction is slightly different for the three considered cases. This is explained by considering that after the application of the proposed \name{} approach, the center of mass of the \gls{ris} shape is changed, and therefore the required pointing direction towards the \gls{ue}.

\subsection{Analysis of the convergence of \name{}}
\begin{figure}
    \centering
    \includegraphics[width=\columnwidth]{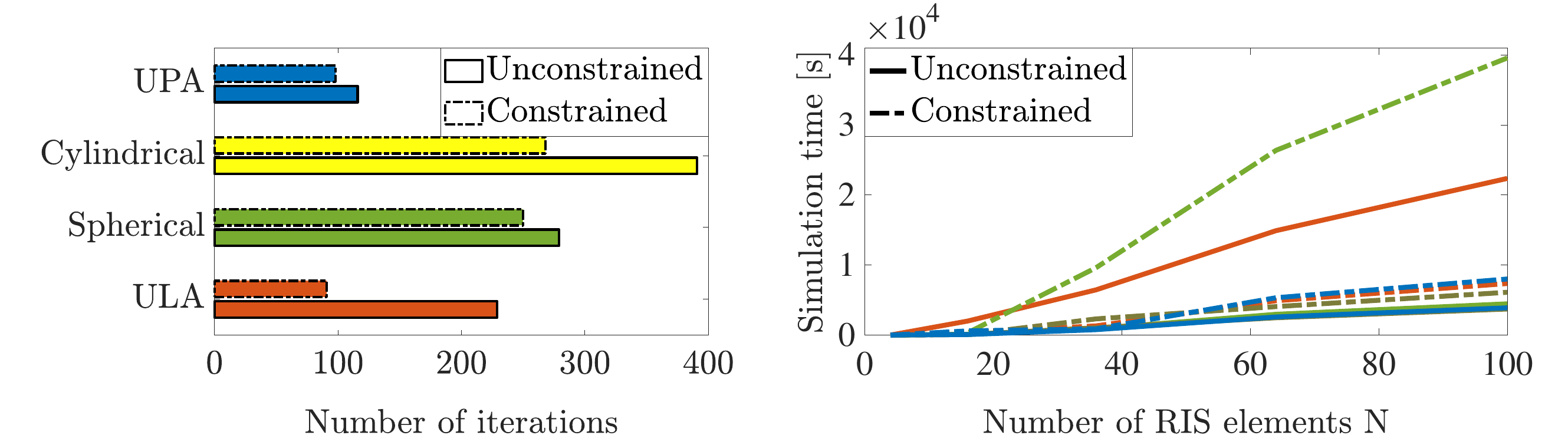}
    \caption{\name{} convergence and simulation time versus the number of RIS elements $N$.}
    \label{fig:convergence}
\end{figure}

In Fig.~\ref{fig:convergence}, on the left-hand side, we empirically show that \name{} is convergent in a finite number of iterations in the considered reference scenarios. We note that the constrained case requires in general fewer iterations, since the feasible space is greatly reduced compared to the unconstrained case. Moreover, on the right-hand side, we show the simulation time versus the number of \gls{ris} elements $N$ for the four considered cases, with color codes given on the left-hand side, thus validating the analysis provided in Sections~\ref{subsec:CA} and~\ref{subsec:CC}.  


\section{Related work}\label{sec:related}

Non-conventionally shaped antennas and arrays are becoming increasingly popular due to their multifunctional capabilities and deployments on supporting surfaces going far beyond traditional flat structures~\cite{mohamadzade2020recent}. A mechanical metasurface constructed from filamentary metal traces that can self-evolve to morph into a wide range of \gls{3d} target shapes is presented in~\cite{Bai_Nature22}. 
In~\cite{ha2012reconfigurable}, a radiating structure of combined loop dipoles is placed on top of a flexible \gls{pcb} substrate, enabling the antenna to be bent and to be adapted to the geometry of the surface of deployment. A flexible patch antenna made by conductive fabric encapsulated in a flexible polymer substrate for wearable devices is presented in~ \cite{simorangkir2017method}. The use of conductive fabric is proposed in~\cite{di2020reconfigurable} and~\cite{salleh2017textile}, where the resulting radiating elements can be directly embedded into clothes. Moreover, the performance of frequency selective curved antenna arrays is assessed in~\cite{yuan2007accurate}. In~\cite{wu2019proactive}, a cylindrical array is proposed to focus the beams in the near field. A flexible antenna array capable of self-adaptation to the surface of deployment is proposed in~\cite{braaten2012self}. Conformal antenna arrays are seen as an aerodynamically efficient solution to deploy large arrays on air and space-borne applications in~\cite{morton2004pattern, nunna2023design}. In~\cite{vellucci2023phase} a cylindrical array of antennas is proposed to obtain $360^\circ$ beamforming.

Recently, non-planar structures are investigated in the \gls{ris} realm. A case study focused on conformal \gls{ris} for \gls{v2v} communication is considered in~\cite{tagliaferri2022conformal}, wherein the performance of an \gls{ris} to be mounted on the side of a car is studied. In particular, the \gls{ris} elements are distributed over a cylindrical surface to be mounted on the side of vehicles and to be configured with a predefined (static) reflection angle. 
A metamorphic smart surface with shape-changing properties is proposed in~\cite{Zelaya21}. Such surfaces can be deployed on objects whose shape changes over time, e.g., on curtains and blinds. Interestingly, it is also demonstrated that non-planar surfaces can match the performance of \glspl{upa} with up to $68\%$ fewer antenna elements.
In~\cite{Nolan21}, the shape of a retro-reflective passive array is mechanically changed to modify the reflection pattern and embed radar-readable information in the reflected signal. In~\cite{Ma23}, the geometry of each unit cell is changed to obtain a frequency-tunable smart surface system. This is achieved by employing antenna elements designed as a rollable thin plastic film with flexible copper strips whose length is tuned to change the surface operating frequency for multi-network coverage extension applications.
The concept of non-planar \glspl{ris} in wearable devices to enable low-power sensing through the modulation of the reflected signal towards the \gls{bs} is envisioned in~\cite{renzo2019smart}.

The interesting properties of non-regularly (conformal) shaped arrays come with a non-negligible mutual coupling among the configurable elements, which has to be taken into account for their optimization and synthesis~\cite{Coe70}. To overcome such a problem, the characterization of the admittance matrix of antenna arrays serves as a valuable optimization proxy that incorporates both the mutual coupling and the configuration of the array elements. Hence, it provides a viable way to synthesize the desired response~\cite{Coe70,Mautz73}.
In~\cite{Qian20}, the admittance matrix is leveraged to model the \gls{e2e} \gls{ris}-aided channel response. In the absence of mutual coupling, this approach leads to a closed-form expression for the optimal \gls{ris} configuration. When mutual coupling is present, iterative optimization algorithms are derived to account for its influence.
In~\cite{Abrardo21}, it is demonstrated that accounting for the mutual coupling during the optimization stage results in better system performance. This idea is pushed to the next stage in~\cite{Mursia23, Hassani23}, wherein the admittance is used to take into account the presence of non-controllable passive scatterers (i.e., the multipath).

While conformal \glspl{ris} are gaining momentum, related studies are limited in terms of arbitrarily shaped structures, and overlook the crucial aspect of accurately modeling the mutual coupling among the elements. To address this fundamental research gap, we present a pioneering optimization framework that enables the joint optimization of the \gls{ris} response and spatial distribution of its elements, while adhering to the constraints imposed by the shape of the deployment support and exploiting the mutual coupling by design. To the best of our knowledge, this is the first work addressing these fundamental aspects of \gls{ris} design in the context of applications and deployments in future smart radio environments.

\section{Conclusions}\label{sec:conclusion}
This paper has investigated the complex reflective properties of conformal surfaces. Our findings unequivocally demonstrate that non-planar \gls{ris} designs and ad-hoc configurations surpass conventional planar approaches in terms of versatility and adaptability, thus heralding a paradigm shift in \gls{ris} optimization.
Our findings demonstrate that such a counter-intuitive integration of non-planar geometries enables the dynamic control over \gls{em} waves in \gls{3d}, opening up new avenues for innovation and optimization in wireless communications. 
In particular, we leverage a recently proposed mutual coupling model for \glspl{ris} to propose \name{}: an iterative algorithm that jointly optimizes the configuration and locations of the \gls{ris} elements in \gls{3d}, subject to realistic deployment and implementation constraints. Our empirical and numerical evaluation using full-wave simulations show the superiority of \name{} with respect to \gls{sota} solutions.
The insights gained from this paper serve as a \emph{foundation for further exploration and advancements in the field of smart surface technologies}.

\begin{appendices}
\section{Self/mutual impedance between RIS elements}
\label{ap:A1}

We start from the closed-form expression of the mutual impedance between any two antennas $p$ and $q$ available in~\cite{DiRenzo23}, by assuming that all the dipole elements have the same half-length $h$
\begin{align}
\nonumber Z_{qp}\! =\! \frac{\eta c}{8\pi} \!\!\sum_{s_o = \{-1, 1 \}} s_0 e^{j s_0 \frac{2\pi}{\lambda} h} \bigg[ I_{qp}(h, s_0) + I_{qp}(-h, s_0)\\
- 2\cos(2\pi h/\lambda)I_{qp}(0, s_0)\bigg],\label{eq:z_qp}
\end{align}
where $c = \frac{1}{\sin(2\pi h/\lambda)}$, and the function $I_{qp}(\xi, s_0)$ is given by
\begin{align}
I_{qp}(\xi,s_0) &=
\mathcal{I}(-s_0, \deltab_{qp}-\xi\ve_3, -h, 0 ) \nonumber \\&+ \mathcal{I}(s_0, \deltab_{qp}-\xi\ve_3, 0, h), \label{eq:I_qp}
\end{align}
where $\deltab_{qp}=\vq_q-\vq_p$, with $\vq_q\in\Real^{3\times 1}$ and $\vq_p\in\Real^{3\times 1}$ denoting the positions of the two considered dipoles. Moreover, the function $\mathcal{I}(s_0,\xib,L,U)$ is defined as
\begin{align}
    \mathcal{I}(s_0,\xib,L,U) =  s_0 e^{j \frac{2\pi}{\lambda} s_0 \ve_3^{\tran}\xib} \big[&T_0(\xib+L\ve_3,s_0)
    \nonumber\\&-T_0(\xib+U\ve_3,s_0)\big], \label{eq:I}
\end{align}
with $T_0(\zetab,s_0)$ given in~\eqref{eq:T}. When setting $h=\lambda/4$, we obtain $c = 1$, $\cos(2\pi h/\lambda) = 0$ and $e^{j\frac{2\pi}{\lambda}s_0h}=s_0j$, which allows us to simplify the expression in \eqref{eq:z_qp} as
\begin{align}
Z_{qp} = \frac{j \eta}{8\pi} \sum_{s_o = \{-1, +1 \}} \big[ &I_{qp}(h, s_0) + I_{qp}(-h, s_0)\big].\label{eq:z_qp_2}
\end{align}
By substituting \eqref{eq:I_qp} and \eqref{eq:I} into \eqref{eq:z_qp_2}, we obtain \eqref{eq:zqp_compact}.

\section{Gradient of $Z_{qp}$ with respect to the position of the RIS elements}\label{ap:A3}

We compute the expression of the gradient of $Z_{qp}$ with respect to the position $\vq_{q}$ and $\vq_{p}$ of the dipoles $q$ and $p$, respectively.
For simplicity, we rewrite  \eqref{eq:zqp_compact} as
\begin{align}
Z_{qp} = \frac{\eta}{8\pi} \sum_{s_o = \{-1, +1 \}} e^{j \frac{2\pi}{\lambda} s_0 \ve_3^{\tran}\deltab}g(\deltab_{qp}, s_0).\label{eq:zqp_compact_112}
\end{align}
with $\deltab_{qp} = \vq_q - \vq_p$, and 
\begin{align}
g(\deltab_{qp}, s_0) =& T_0(\deltab_{qp}-2h\ve_3, s_0)+T_0(\deltab_{qp}+2h\ve_3, s_0)\nonumber\\&-2T_0(\deltab_{qp}, s_0). \label{eq:g}
\end{align}
The gradient of \eqref{eq:zqp_compact_112} with respect to $\deltab_{qp}$ is
\begin{align}
\nonumber \nabla Z_{qp} = \frac{\eta}{8\pi} \!\!\! \sum_{s_o = \{-1, 1 \}}\!\!\! \Big( j \frac{2\pi}{\lambda} s_0 \ve_3 e^{j \frac{2\pi}{\lambda} s_0 \ve_3^{\tran}\deltab_{qp}}g(\deltab_{qp}, s_0)\\ 
+ e^{j \frac{2\pi}{\lambda} s_0 \ve_3^{\tran}\deltab_{qp}}\nabla g(\deltab_{qp}, s_0) \Big),\label{eq:dzqp_compact_1}
\end{align}
with
\begin{align}
\nabla g(\deltab_{qp}, s_0) = &
\nabla T_0(\deltab_{qp}-2h\ve_3, s_0)+\nabla T_0(\deltab_{qp}+2h\ve_3, s_0)
\nonumber\\&-2\nabla T_0(\deltab_{qp}, s_0). \label{eq:dg_ddelta}
\end{align}
Since $\frac{d\, E_1(x)}{d\, x}=-\frac{e^{-x}}{x}$, with \change{$E_1(x) = \int_{x}^\infty \frac{e^{-t}}{t} dt$} the exponential integral, we have that
\begin{align}
\nabla T_0(\zetab,s_0) = -\Big(\frac{ \zetab}{\|\zetab\|}+s_0\ve_3\Big) \frac{e^{-j\frac{2\pi}{\lambda}(\|\zetab \|+s_0\ve_3^{\tran} \zetab)}}{\|\zetab \|+s_0\ve_3^{\tran} \zetab}.\label{eq:dT}
\end{align}
Moreover, we have that
\change{
$\nabla \deltab_{qp}^{\tran} = \mI_3$, and
$\nabla \deltab_{qp}^{\tran} = -\mI_3$.
}
We are now in the position of obtaining $\nabla Z_{qp}$ by substituting \eqref{eq:dg_ddelta}-\eqref{eq:dT} in \eqref{eq:dzqp_compact_1}, which gives
\begin{align}
\nonumber\nabla Z_{qp}\!\! = & \frac{\eta}{8\pi} \!\!\! \sum_{s_o = \{-1, 1 \}} \!\!\!\!\! \Big( j \frac{2\pi}{\lambda} s_0 \ve_3 e^{j \frac{2\pi}{\lambda} s_0 \ve_3^{\tran}(\vq_q-\vq_p)}g(\vq_q-\vq_p, s_0)\\  
&+ e^{j \frac{2\pi}{\lambda} s_0 \ve_3^{\tran}(\vq_q-\vq_p)}\nabla g(\vq_q-\vq_p, s_0) \Big) \label{eq:grad_Z_q}.
\end{align}
\end{appendices}

\bibliographystyle{IEEEtran}
\bibliography{IEEEabrv, references}

\vfill

\end{document}

%% file: Definitions.tex
\makeatletter
\let\oldabs\abs
\def\abs{\@ifstar{\oldabs}{\oldabs*}}

\let\oldnorm\norm
\def\norm{\@ifstar{\oldnorm}{\oldnorm*}}
\makeatother

\DeclareMathOperator{\vb}{\mathbf{b}}

\DeclareMathOperator{\ve}{\mathbf{e}}

\DeclareMathOperator{\vg}{\mathbf{g}}

\DeclareMathOperator{\vp}{\mathbf{p}}
\DeclareMathOperator{\vq}{\mathbf{q}}

\DeclareMathOperator{\vx}{\mathbf{x}}

\DeclareMathOperator{\vz}{\mathbf{z}}

\DeclareMathOperator{\mG}{\mathbf{G}}

\DeclareMathOperator{\mI}{\mathbf{I}}

\DeclareMathOperator{\mQ}{\mathbf{Q}}

\DeclareMathOperator{\mZ}{\mathbf{Z}}

\DeclareMathOperator{\deltab}{\boldsymbol{\delta}}

\DeclareMathOperator{\zetab}{\boldsymbol{\zeta}}

\DeclareMathOperator{\xib}{\boldsymbol{\xi}}

\DeclareMathOperator{\rhob}{\boldsymbol{\rho}}

\DeclareMathOperator{\Gammab}{\mathbf{\Gamma}}
\DeclareMathOperator{\Deltab}{\mathbf{\Delta}}

\DeclareMathOperator{\Real}{\mbox{$\mathbb{R}$}}
\DeclareMathOperator{\Compl}{\mbox{$\mathbb{C}$}}

\DeclareMathOperator{\Exp}{\mathbb{E}}

\DeclareMathOperator{\herm}{\mathrm{H}}

\DeclareMathOperator{\tr}{\mathrm{tr}}
\DeclareMathOperator{\tran}{\mathrm{T}}

%% file: main.bbl
\begin{thebibliography}{10}
\providecommand{\url}[1]{#1}
\csname url@samestyle\endcsname
\providecommand{\newblock}{\relax}
\providecommand{\bibinfo}[2]{#2}
\providecommand{\BIBentrySTDinterwordspacing}{\spaceskip=0pt\relax}
\providecommand{\BIBentryALTinterwordstretchfactor}{4}
\providecommand{\BIBentryALTinterwordspacing}{\spaceskip=\fontdimen2\font plus
\BIBentryALTinterwordstretchfactor\fontdimen3\font minus
  \fontdimen4\font\relax}
\providecommand{\BIBforeignlanguage}[2]{{%
\expandafter\ifx\csname l@#1\endcsname\relax
\typeout{** WARNING: IEEEtran.bst: No hyphenation pattern has been}%
\typeout{** loaded for the language `#1'. Using the pattern for}%
\typeout{** the default language instead.}%
\else
\language=\csname l@#1\endcsname
\fi
#2}}
\providecommand{\BIBdecl}{\relax}
\BIBdecl

\bibitem{DiRenzo2020_Jsac}
M.~Di~Renzo, A.~Zappone \emph{et~al.}, ``Smart radio environments empowered by
  reconfigurable intelligent surfaces: How it works, state of research, and the
  road ahead,'' \emph{{IEEE} J. Sel. Areas Commun.}, vol.~38, no.~11, pp.
  2450--2525, 2020.

\bibitem{long2021promising}
W.~Long, R.~Chen \emph{et~al.}, ``A promising technology for 6g wireless
  networks: Intelligent reflecting surface,'' \emph{J. Commun. Informat.
  Netw.}, vol.~6, no.~1, pp. 1--16, 2021.

\bibitem{renzo2019smart}
M.~Di~Renzo, M.~Debbah \emph{et~al.}, ``Smart radio environments empowered by
  reconfigurable ai meta-surfaces: An idea whose time has come,''
  \emph{{EURASIP} J. Wireless Commun. Netw.}, vol. 2019, no.~1, pp. 1--20,
  2019.

\bibitem{basharat2021reconfigurable}
S.~Basharat, S.~A. Hassan \emph{et~al.}, ``Reconfigurable intelligent surfaces:
  Potentials, applications, and challenges for 6g wireless networks,''
  \emph{{IEEE} Wireless Commun.}, vol.~28, no.~6, pp. 184--191, 2021.

\bibitem{Cui2023}
Z.~Cui and S.~Pollin, ``Impact of reconfigurable intelligent surface geometry
  on communication performance,'' \emph{{IEEE} Wireless Commun. Lett.},
  vol.~12, no.~5, pp. 898--902, 2023.

\bibitem{Lin2022}
Y.~Lin, S.~Jin \emph{et~al.}, ``Conformal irs-empowered mimo-ofdm: Channel
  estimation and environment mapping,'' \emph{{IEEE} Trans. Commun.}, vol.~70,
  no.~7, pp. 4884--4899, 2022.

\bibitem{Budhu22}
J.~Budhu, L.~Szymanski, and A.~Grbic, ``Design of planar and conformal,
  passive, lossless metasurfaces that beamform,'' \emph{IEEE J. Microw.},
  vol.~2, no.~3, pp. 401--418, 2022.

\bibitem{Mizmizi23}
M.~Mizmizi, R.~A. Ayoubi \emph{et~al.}, ``Conformal metasurfaces: A novel
  solution for vehicular communications,'' \emph{{IEEE} Trans. Wireless
  Commun.}, vol.~22, no.~4, pp. 2804--2817, 2023.

\bibitem{Zelaya21}
R.~I. Zelaya, R.~Ma, and W.~Hu, ``{Towards 6G and Beyond: Smarten Everything
  with Metamorphic Surfaces},'' in \emph{Proc. 20th ACM HotNets Workshop},
  2021, pp. 155--162.

\bibitem{Gradoni2021}
G.~Gradoni and M.~{Di Renzo}, ``{End-to-End Mutual Coupling Aware Communication
  Model for Reconfigurable Intelligent Surfaces: An Electromagnetic-Compliant
  Approach Based on Mutual Impedances},'' \emph{{IEEE} Wireless Commun. Lett.},
  vol. 2337, no.~c, pp. 1--5, 2021.

\bibitem{MarcoSergei2022}
M.~Di~Renzo, F.~H. Danufane, and S.~Tretyakov, ``Communication models for
  reconfigurable intelligent surfaces: From surface electromagnetics to
  wireless networks optimization,'' \emph{Proc. of the IEEE}, vol. 110, no.~9,
  pp. 1164--1209, 2022.

\bibitem{Nolan21}
J.~Nolan, K.~Qian, and X.~Zhang, ``{RoS: Passive smart surface for
  roadside-to-vehicle communication},'' in \emph{Proc. ACM SIGCOMM}, 2021, pp.
  165--178.

\bibitem{Ma23}
R.~Ma, R.~I. Zelaya, and W.~Hu, ``{Softly, Deftly, Scrolls Unfurl Their
  Splendor: Rolling Flexible Surfaces for Wideband Wireless},'' in \emph{Proc.
  ACM Mobicom}, 2023, pp. 1--15.

\bibitem{SmartSkins}
G.~Oliveri, P.~Rocca \emph{et~al.}, ``Holographic smart {EM} skins for advanced
  beam power shaping in next generation wireless environments,'' \emph{IEEE J.
  Multiscale Multiphysics Comput. Techniques}, vol.~6, pp. 171--182, 2021.

\bibitem{FluidAntennas}
K.-K. Wong, W.~K. New \emph{et~al.}, ``Fluid antenna system—{Part I}:
  Preliminaries,'' \emph{IEEE Commun. Lett.}, vol.~27, no.~8, pp. 1919--1923,
  2023.

\bibitem{MovableAntennas}
L.~Zhu, W.~Ma, and R.~Zhang, ``Modeling and performance analysis for movable
  antenna enabled wireless communications,'' arXiv, 2210.05325, 2022.

\bibitem{mattiello2015analysis}
F.~Mattiello, G.~Leone, and G.~Ruvio, ``Analysis and characterization of
  finite-size curved frequency selective surfaces,'' \emph{Studies in
  Engineering and Technology}, vol.~2, no.~1, 2015.

\bibitem{chaumet2022discrete}
P.~C. Chaumet, ``The discrete dipole approximation: A review,''
  \emph{Mathematics}, vol.~10, no.~17, p. 3049, 2022.

\bibitem{gibson2007method}
W.~Gibson, \emph{The Method of Moments in Electromagnetics}.\hskip 1em plus
  0.5em minus 0.4em\relax CRC Press, 2007.

\bibitem{zhang2021hyperuniform}
H.~Zhang, Q.~Cheng \emph{et~al.}, ``Hyperuniform disordered distribution
  metasurface for scattering reduction,'' \emph{Applied Physics Lett.}, vol.
  118, no.~10, 2021.

\bibitem{yang2019surface}
F.~Yang and Y.~Rahmat-Samii, \emph{Surface Electromagnetics: With Applications
  in Antenna, Microwave, and Optical Engineering}.\hskip 1em plus 0.5em minus
  0.4em\relax Cambridge University Press, 2019.

\bibitem{Bosiljevac2020}
D.~Barbarić, M.~Bosiljevac, and Z.~Šipuš, ``Analysis of curved metasurfaces
  based on method of moments,'' in \emph{Proc. European Conf. Antennas
  Propag.}, 2020, pp. 1--5.

\bibitem{Craeye2019}
M.~Bodehou, D.~González-Ovejero \emph{et~al.}, ``Method of moments simulation
  of modulated metasurface antennas with a set of orthogonal entire-domain
  basis functions,'' \emph{{IEEE} Trans. Antennas Propag.}, vol.~67, no.~2, pp.
  1119--1130, 2019.

\bibitem{konno2016fast}
K.~Konno, Q.~Chen, and R.~J. Burkholder, ``Fast computation of layered media
  green's function via recursive taylor expansion,'' \emph{IEEE Antennas and
  Wireless Propag. Lett.}, vol.~16, pp. 1048--1051, 2016.

\bibitem{konno2023generalised}
K.~Konno, S.~Terranova \emph{et~al.}, ``Generalised impedance model of wireless
  links assisted by reconfigurable intelligent surfaces,'' 2023.

\bibitem{mursia2023empirical}
P.~Mursia, T.~Mazloum \emph{et~al.}, ``Empirical validation of the
  impedance-based ris channel model in an indoor scattering environment,'' in
  \emph{Proc. European Conf. Antennas Propag.}, 2023.

\bibitem{zheng2024mutual}
P.~Zheng, R.~Wang \emph{et~al.}, ``Mutual coupling in ris-aided communication:
  Model training and experimental validation,'' 2024.

\bibitem{Vukovic2022}
T.~Dimitrijevic, J.~Jokovic \emph{et~al.}, ``Tlm mesh impact on textile antenna
  modelling,'' in \emph{Proc. Microwave Mediterranean Symposium}, 2022, pp.
  1--4.

\bibitem{STAR-RIS}
Y.~Liu, X.~Mu \emph{et~al.}, ``{STAR}: Simultaneous transmission and reflection
  for 360° coverage by intelligent surfaces,'' \emph{{IEEE} Wireless Commun.},
  vol.~28, no.~6, pp. 102--109, 2021.

\bibitem{OMNI-RIS}
H.~Zhang, S.~Zeng \emph{et~al.}, ``Intelligent omni-surfaces for
  full-dimensional wireless communications: Principles, technology, and
  implementation,'' \emph{{IEEE} Commun. Mag.}, vol.~60, no.~2, pp. 39--45,
  2022.

\bibitem{sipus2007analysis}
Z.~Sipus, M.~Bosiljevac, and S.~Skokic, ``Analysis of curved frequency
  selective surfaces,'' in \emph{Proc. European Conf. Antennas Propag.}, 2007,
  pp. 1--5.

\bibitem{DiRenzo23}
M.~Di~Renzo, V.~Galdi, and G.~Castaldi, ``Modeling the mutual coupling of
  reconfigurable metasurfaces,'' in \emph{Proc. European Conf. Antennas
  Propag.}, 2023, pp. 1--4.

\bibitem{Mursia23}
P.~Mursia, S.~Phang \emph{et~al.}, ``{SARIS}: Scattering aware reconfigurable
  intelligent surface model and optimization for complex propagation
  channels,'' \emph{{IEEE} Wireless Commun. Lett.}, vol.~12, no.~11, pp. 1--1,
  2023.

\bibitem{Hassani23}
\BIBentryALTinterwordspacing
H.~E. Hassani, X.~Qian \emph{et~al.}, ``{Optimization of RIS-Aided MIMO -- A
  Mutually Coupled Loaded Wire Dipole Model},'' 2023. [Online]. Available:
  \url{http://arxiv.org/abs/2306.09480}
\BIBentrySTDinterwordspacing

\bibitem{Qian20}
X.~Qian and M.~{Di Renzo}, ``{Mutual coupling and unit cell aware optimization
  for reconfigurable intelligent surfaces},'' \emph{{IEEE} Wireless Commun.
  Lett.}, vol.~10, no.~6, pp. 10--14, 2020.

\bibitem{Abrardo21}
A.~Abrardo, D.~Dardari \emph{et~al.}, ``{MIMO Interference Channels Assisted by
  Reconfigurable Intelligent Surfaces: Mutual Coupling Aware Sum-Rate
  Optimization Based on a Mutual Impedance Channel Model},'' \emph{{IEEE}
  Wireless Commun. Lett.}, vol.~10, no.~12, pp. 2624--2628, 2021.

\bibitem{Bertsekas}
D.~P. Bertsekas, \emph{Nonlinear programming}.\hskip 1em plus 0.5em minus
  0.4em\relax Athena Scientific, 1999.

\bibitem{cststudio}
{Dassault Syst\`{e}mes}, ``{CST Studio Suite: Electromagnetic field simulation
  software},''
  \url{https://www.3ds.com/products-services/simulia/products/cst-studio-suite/},
  2022.

\bibitem{Russer2023}
J.~A. Russer, D.~Semmler \emph{et~al.}, ``Analysis of intelligent surface-aided
  mimo communication systems,'' in \emph{Proc. 26th Int. ITG WSA \& SCC}, 2023,
  pp. 1--5.

\bibitem{nerini2023universal}
M.~Nerini, S.~Shen \emph{et~al.}, ``A universal framework for multiport network
  analysis of reconfigurable intelligent surfaces,'' 2023.

\bibitem{abrardo2023design}
A.~Abrardo, A.~Toccafondi, and M.~D. Renzo, ``Design of reconfigurable
  intelligent surfaces by using s-parameter multiport network theory --
  optimization and full-wave validation,'' 2023.

\bibitem{Pozar}
D.~M. Pozar, \emph{{Microwave Engineering}}.\hskip 1em plus 0.5em minus
  0.4em\relax Wiley Press, 2011.

\bibitem{mohamadzade2020recent}
B.~Mohamadzade, R.~B. Simorangkir \emph{et~al.}, ``Recent developments and
  state of the art in flexible and conformal reconfigurable antennas,''
  \emph{Electronics}, vol.~9, no.~9, p. 1375, 2020.

\bibitem{Bai_Nature22}
Y.~Bai, H.~Wang \emph{et~al.}, ``A dynamically reprogrammable metasurface with
  self-evolving shape morphing,'' \emph{Nature}, vol. 609, pp. 701--708, 2022.

\bibitem{ha2012reconfigurable}
S.-J. Ha, Y.-B. Jung \emph{et~al.}, ``Reconfigurable beam-steering antenna
  using dipole and loop combined structure for wearable applications,''
  \emph{ETRI journal}, vol.~34, no.~1, pp. 1--8, 2012.

\bibitem{simorangkir2017method}
R.~B. Simorangkir, Y.~Yang \emph{et~al.}, ``A method to realize robust flexible
  electronically tunable antennas using polymer-embedded conductive fabric,''
  \emph{{IEEE} Trans. Antennas Propag.}, vol.~66, no.~1, pp. 50--58, 2017.

\bibitem{di2020reconfigurable}
A.~Di~Natale and E.~Di~Giampaolo, ``A reconfigurable all-textile wearable uwb
  antenna,'' \emph{Progress In Electromagnetics Research C}, vol. 103, pp.
  31--43, 2020.

\bibitem{salleh2017textile}
S.~M. Salleh, M.~Jusoh \emph{et~al.}, ``Textile antenna with simultaneous
  frequency and polarization reconfiguration for wban,'' \emph{IEEE Access},
  vol.~6, pp. 7350--7358, 2017.

\bibitem{yuan2007accurate}
N.~Yuan, X.-C. Nie \emph{et~al.}, ``Accurate analysis of conformal antenna
  arrays with finite and curved frequency selective surfaces,'' \emph{J.
  Electromagn. Waves Appl.}, vol.~21, no.~13, pp. 1745--1760, 2007.

\bibitem{wu2019proactive}
Y.~F. Wu and Y.~J. Cheng, ``Proactive conformal antenna array for near-field
  beam focusing and steering based on curved substrate integrated waveguide,''
  \emph{{IEEE} Trans. Antennas Propag.}, vol.~67, no.~4, pp. 2354--2363, 2019.

\bibitem{braaten2012self}
B.~D. Braaten, S.~Roy \emph{et~al.}, ``A self-adapting flexible (selflex)
  antenna array for changing conformal surface applications,'' \emph{{IEEE}
  Trans. Antennas Propag.}, vol.~61, no.~2, pp. 655--665, 2012.

\bibitem{morton2004pattern}
T.~E. Morton and K.~M. Pasala, ``Pattern synthesis of conformal arrays for
  airborne vehicles,'' in \emph{Proc. IEEE Aerospace Conf.}, vol.~2.\hskip 1em
  plus 0.5em minus 0.4em\relax IEEE, 2004, pp. 1030--1039.

\bibitem{nunna2023design}
B.~A. Nunna and V.~K. Kothapudi, ``Design and analysis of x-band conformal
  antenna array for spaceborne synthetic aperture radar applications,'' in
  \emph{Proc. Second Int. Conf. Computational Electron. Wireless Commun.}\hskip
  1em plus 0.5em minus 0.4em\relax Springer, 2023, pp. 193--204.

\bibitem{vellucci2023phase}
S.~Vellucci, M.~Longhi \emph{et~al.}, ``Phase-gradient huygens metasurface
  coatings for dynamic beamforming in linear antennas,'' 2023.

\bibitem{tagliaferri2022conformal}
D.~Tagliaferri, M.~Mizmizi \emph{et~al.}, ``Conformal intelligent reflecting
  surfaces for 6g v2v communications,'' in \emph{Proc. 1st Int. Conf. 6G
  Netw.}\hskip 1em plus 0.5em minus 0.4em\relax IEEE, 2022, pp. 1--8.

\bibitem{Coe70}
R.~Coe and A.~Ishimaru, ``Optimum scattering from an array of half-wave
  dipoles,'' \emph{{IEEE} Trans. Antennas Propag.}, vol.~18, no.~2, pp.
  224--230, 1970.

\bibitem{Mautz73}
J.~Mautz and R.~Harrington, ``Modal analysis of loaded n-port scatterers,''
  \emph{{IEEE} Trans. Antennas Propag.}, vol.~21, no.~2, pp. 188--199, 1973.

\end{thebibliography}
